\documentclass{aa} 

\usepackage{graphicx}
\usepackage{txfonts}

\usepackage[english]{babel}
\usepackage{amsmath}    
\usepackage{amssymb}    
\usepackage{multicol}        
\usepackage{bm}         
\usepackage{pdflscape}  
\usepackage{flafter}
\usepackage{wrapfig}
\usepackage[export]{adjustbox}
\usepackage{float}
\usepackage{enumitem}
\usepackage{tabularx} 
\usepackage[normalem]{ulem}
\usepackage{subfig}
\usepackage{tablefootnote}

\usepackage[table,xcdraw]{xcolor}
\usepackage{afterpage}

\usepackage[colorlinks=true,linkcolor=blue,citecolor=blue,urlcolor=blue]{hyperref}

\begin{document} 
   
   \title{Probing Na in giant exoplanets with ESPRESSO and 3D NLTE stellar spectra\thanks{The grid of 3D NLTE synthetic spectra for FGK-type dwarfs is only available in electronic form at \url{https://doi.org/10.5281/zenodo.13881828.}}}

   \author{G.~Canocchi\inst{\ref{inst1}} \and
          G. ~Morello\inst{\ref{inst2}, \ref{inst3}} \and
          K.~Lind\inst{\ref{inst1}} \and
          I. ~Carleo\inst{\ref{inst4}, \ref{inst3}, \ref{inst5}} \and
          M. ~Stangret\inst{\ref{inst6}} \and
          E. ~Pall\'{e}\inst{\ref{inst3}, \ref{inst5}}
          }
     \institute{Department of Astronomy, Stockholm University, AlbaNova University Center, SE-106 91 Stockholm, Sweden\\ \email{gloria.canocchi@astro.su.se}\label{inst1} \and 
    Instituto de Astrof\'{i}sica de Andaluc\'{i}a (IAA-CSIC), Gta. de la Astronom\'{i}a s/n, 18008 Granada, Granada, Spain\label{inst2} \and
    Instituto de Astrof\'{i}sica de Canarias (IAC), E-38205 La Laguna, Tenerife, Spain\label{inst3} \and
    INAF – Osservatorio Astrofisico di Torino, Via Osservatorio 20, I-10025, Pino Torinese, Italy\label{inst4} \and
    Departamento de Astrof\'{i}sica, Universidad de La Laguna (ULL), E-38206 La Laguna, Tenerife, Spain\label{inst5} \and
    INAF – Osservatorio Astronomico di Padova, Vicolo dell’Osservatorio 5, 35122, Padova, Italy\label{inst6}
    }

 
  \abstract  
   {Neutral sodium was the first atom detected in an exoplanetary atmosphere using the transmission spectroscopy technique. To date, it remains the most successfully detected species due to its strong doublet in the optical at 5890\,\r{A} and 5896\,\r{A}. However, the center-to-limb variation (CLV) of these lines in the host star can bias the \ion{Na}{I} detection. In fact, when combined with the Rossiter-McLaughlin (RM) effect, the CLV can mimic or obscure a planetary absorption feature if not properly accounted for.} 
   {This work aims to investigate the impact of three-dimensional (3D) radiation hydrodynamic stellar atmospheres and non-local thermodynamic equilibrium (NLTE) radiative transfer on the modeling of the CLV+RM effect in single-line transmission spectroscopy, to improve the detection and characterization of exoplanet atmospheres.}
   {
   We produced a grid of 3D non-LTE synthetic spectra for \ion{Na}{I} for FGK-type dwarfs within the following parameter space: $T_\mathrm{eff}=4500-6500$\,K, $\log{g} =4.0-5.0$ and [Fe/H]= [-0.5, 0, 0.5]. This grid was then interpolated to match the stellar parameters of four stars hosting well-known giant exoplanets, generating stellar spectra to correct for the CLV+RM effect in their transmission spectra. We used archival observations taken with the high-resolution ESPRESSO spectrograph.  
    }
   {
   Our work confirms the \ion{Na}{I} detections in three systems, namely WASP-52b, WASP-76b, and WASP-127b, improving the accuracy of the measured absorption depth. Furthermore, we find that 3D NLTE stellar models can explain the spectral features in the transmission spectra of HD 209458b without the need for any planetary absorption.
   In the grid of stellar synthetic spectra, we observe that the CLV effect is stronger for stars with low $T_\mathrm{eff}$ and high $\log g$. 
   However, the combined effect of CLV and RM is highly dependent on the orbital geometry of the planet-star system.
   } 
   {With the continuous improvement of instrumentation, it is crucial to use the most accurate stellar models available to correct for the CLV+RM effect in high-resolution transmission spectra, to achieve the best possible characterization of exoplanet atmospheres. This will be fundamental in preparation for instruments such as ANDES at the Extremely Large Telescope, to fully exploit its capabilities in the near future. We make our grid of 3D NLTE synthetic spectra for \ion{Na}{I} publicly available.}

\titlerunning{3D NLTE modeling of the stellar CLV in gas giants}
\authorrunning{Canocchi G. et al.}
   \maketitle

\section{Introduction}\label{sec: intro}
The study of the composition and dynamics of the atmospheres of giant exoplanets has seen enormous progress in the last two decades, starting in 2002 from the first detection of \ion{Na}{I} in the atmosphere of the Hot Jupiter HD 209458b (\citealt{Charbonneau2002}), through the analysis of low-resolution spectra from the Hubble Space Telescope (HST) via transmission spectroscopy. Only a few years later, the same technique started to be performed from ground-based high-resolution ($R \sim 100 \ 000$) spectrographs as well (e.g., \citealt{Narita2005}; \citealt{Snellen2008}; \citealt{Redfield2008}), making it a powerful and complementary tool to the space telescopes for the analysis of exoplanets atmospheres. Indeed, ground-based spectrographs such as the High Accuracy Radial velocity Planet Searcher in the southern (HARPS; \citealt{Mayor2003}) and northern hemisphere (HARPS-N; \citealt{Cosentino2012}), as well as the Echelle Spectrograph for Rocky Exoplanet and Stable Spectroscopic Observation (ESPRESSO; \citealt{Pepe2010, Pepe2014, Pepe2021}) at the Very Large Telescope (VLT), are able to resolve the spectral line profiles of several strong lines of different atomic species, especially alkali metals such as neutral sodium (\ion{Na}{I}) and potassium (\ion{K}{I}), both of which have strong optical absorption features in stars, as well as in giant planet atmospheres.  
In particular, absorption line profiles of the \ion{Na}{I} doublet at 5890\,\r{A} and 5896\,\r{A} can be spectrally resolved through the analysis of high-resolution transmission spectra to infer several physical quantities such as the temperatures, number densities, and wind patterns at the probed atmospheric layers (e.g., \citealt{Wyttenbach2015}; \citealt{Seidel2019}). 
Apart from \ion{Na}{I}, more than 50 other atomic and molecular species have been detected in more than 200 exoplanets' atmospheres\footnote{\url{https://research.iac.es/proyecto/exoatmospheres/index.php}}. 
The analysis of the composition of exoplanets plays a critical role in understanding and constraining planet formation and evolution theories (e.g., \citealt{Oberg2011}; \citealt{Madhusudhan2014}; \citealt{Mordasini2016}), and transmission spectroscopy is currently the most effective method for atmospheric characterization. 

The technique requires that a time series of spectra be taken before, after, and during the transit of an exoplanet (\citealt{Brown2001}). Indeed, when the planet passes in front of its host star during the primary eclipse, part of the stellar light filters through the upper layers of the planet's atmosphere, and there it is absorbed by the atoms and molecules that are present in it. 
Then, by comparing the spectra taken during the transit event (in-transit) to those taken before and after (out-of-transit), it is possible to recover the planet's signature at specific wavelengths corresponding to the chemical species we are looking for.

An important effect to consider when analyzing transmission spectra with this technique is the center-to-limb variation (CLV) of the stellar line profiles. The CLV describes the variation in strength and shape of the lines forming in the stellar atmosphere as the observer's line of sight moves from the center to the edge of the stellar disk, since lines are forming in increasingly higher layers towards the edge. 
Several previous studies stressed the importance of correctly accounting for this effect on the interpretation of the spectra of gas giants, showing that different stellar models lead to very different results (e.g., \citealt{Yan2017}; \citealt{Borsa2018}; \citealt{Chiavassa2019}; \citealt{Chen2020}).

So far, to model this effect, most of the transmission spectroscopy studies made use of synthetic stellar spectra produced with one-dimensional (1D) plane-parallel stellar atmospheres and Local Thermodynamic Equilibrium (LTE) radiative transfer codes, such as the Spectroscopy Made Easy (SME; \citealt{Valenti1996}; \citealt{Piskunov2017}) tool. In the LTE assumption, atomic level populations are determined by collisions and are thus easily computed using the Saha-Boltzmann equations. LTE is valid in high-density environments, such as the deeper layers of a star, but breaks down in the outer, less dense layers (photosphere) where spectral lines form. In these regions, radiative rates dominate over collisional rates, making the LTE assumption invalid. To accurately model spectral lines in these conditions, the more complex and computationally demanding Non-Local Thermodynamic Equilibrium (NLTE) assumption is required, where the level populations are governed by the statistical equilibrium equations (e.g., \citealt{Rutten2003}). Implementing NLTE modeling requires not only significantly more computational power but also highly accurate atomic and molecular data. Thanks to advancements in computational power, it is now feasible to generate synthetic stellar spectra not only using 1D atmospheres but even with the more accurate three-dimensional (3D) radiation-hydrodynamic (RHD) simulations.
For an in-depth discussion on how different assumptions regarding the model atmosphere (1D/3D) and radiative transfer (LTE/NLTE) affect the stellar synthetic spectra computations, we refer the reader to the recent review by \citet{LindAmarsi2024} and the references therein.

The Solar CLVs of spectral lines are a probe of coupled 3D and NLTE effects since they are mapping different layers of the atmosphere, as mentioned above.
Even if widely used, 1D LTE models fail to reproduce the CLV of several atomic lines in the Sun (e.g., \citealt{Pereira2013}; \citealt{Lind2017}; \citealt{Nordlander2017}; \citealt{Bjorgen2018}; \citealt{Amarsi2018, Amarsi2019}; \citealt{Bergemann2021}; \citealt{pietrow2023centertolimb, Pietrow2023c}), including the \ion{Na}{I} D resonance lines which are affected by strong NLTE effects.  
Some improvements were made by \citet{Yan2017}, who used 1D NLTE spectra of the \ion{Na}{I} doublet to correct for the CLV effect in high-resolution spectra of HD 189733 b. Further progress was achieved by \citet{Reiners2023}, who investigated the impact of the CLV in transmission light curves (LCs) by simulating the Sun as an exoplanet host star. They compared the recently published Solar Atlas (\citealt{Ellwarth2023}) with synthetic spectra computed using 3D RHD atmospheres under the assumption of LTE.  

Recently, \citet{Canocchi2024} showed that stellar spectra produced with 3D model atmospheres and NLTE line formation are required to correctly reproduce the CLV of high-resolution spatially resolved observations of the \ion{Na}{I} D lines at different viewing angles ($\mu$-angles\footnote{$\mu= \cos \theta$, with $\theta$ defined as the angle between the outward normal of the stellar surface and the observer's line of sight.}) in the Solar disk from the Swedish 1-m Solar Telescope (SST; \citealt{Scharmer2003}). The LTE models, whether using 1D or 3D atmospheres, significantly underestimate the strength of the \ion{Na}{I} D lines observed at the Solar limb. These lines are strongly affected by NLTE effects and are highly sensitive to the velocity fields present in the stellar atmosphere. As a result, the traditional 1D plane-parallel model fails to accurately reproduce their CLV, even when NLTE calculations are employed.

Another effect that deforms the shape of the stellar lines as the planet transits the stellar disk is the Rossiter-McLaughlin (RM; \citealt{Rossiter1924}; \citealt{McLaughlin1924}) effect. Since the star is rotating during the transit, and the planet disk is selectively obscuring different parts of the stellar disk, the average wavelength of the stellar line is therefore slightly shifted toward the red or blue in different orbital phases.
The RM effect can be as critical as the CLV effect, and therefore it must be modeled carefully in transmission spectroscopy studies (e.g., \citealt{Czesla2015}; \citealt{Casasayas2020}; \citealt{Casasayas2021}; \citealt{Morello2022})..

Currently, ESPRESSO stands out as the best ground-based facility for analyzing stellar spectra in the optical wavelength range (\citealt{Casasayas2021}), employing transit observations to characterize the atmosphere of exoplanets orbiting around them.
Indeed, it is an ultra-stable spectrograph, with remarkably high spectral and temporal resolution.
In this work, we compute a grid of synthetic stellar spectra for the \ion{Na}{I} D lines using 3D radiation hydrodynamic atmospheres and NLTE radiative transfer for FGK-type dwarfs within the following parameter space: effective temperature ($T_\mathrm{eff}$) between 4500 and 6500\,K, surface gravity ($\log g$) between 4.0 and 5.0\,dex, and metallicity ([Fe/H]) between $-0.5$ and $+0.5$\,dex. It is important to highlight that 3D NLTE synthetic spectra for \ion{Na}{I} did not previously exist for these stars.
We then interpolate the grid to the specific stellar parameters of four stars hosting well-known close-in giant exoplanets, ranging from Warm Neptunes to Ultra Hot Jupiters, to obtain accurate 3D NLTE synthetic spectra.
We then calculate the CLV and RM model from these spectra, and use them to correct for these effects the transmission spectra extracted from archival observations taken with ESPRESSO in the previous years. A comparison with the models computed via 1D LTE, 3D LTE, and 1D NLTE is included in order to test the impact of the 3D NLTE models in improving the (non-)detection of \ion{Na}{I} in the exoplanets' atmosphere.

The paper is organized as follows. Section \ref{sec: obs} gives an overview of the ESPRESSO observations used in this work as well as a brief introduction to each planet-star system. The synthesis of stellar spectra with 3D stellar atmospheres and NLTE radiative transfer for the correction of the CLV effect is explained in Sect. \ref{sec: synthspectra}. The data analysis steps to get the transmission spectra are described in detail in Sect. \ref{sec: methods}. 
Then, in Sect. \ref{sec: analysis}, our results for \ion{Na}{I} detections are analyzed and discussed with reference to previous studies. Finally, in Sect. \ref{sec: conclusions} the main conclusions are summarized and future perspectives are presented.

\section{Observations}\label{sec: obs}
\begin{figure}
    \centering
    \includegraphics[width=0.5\textwidth]{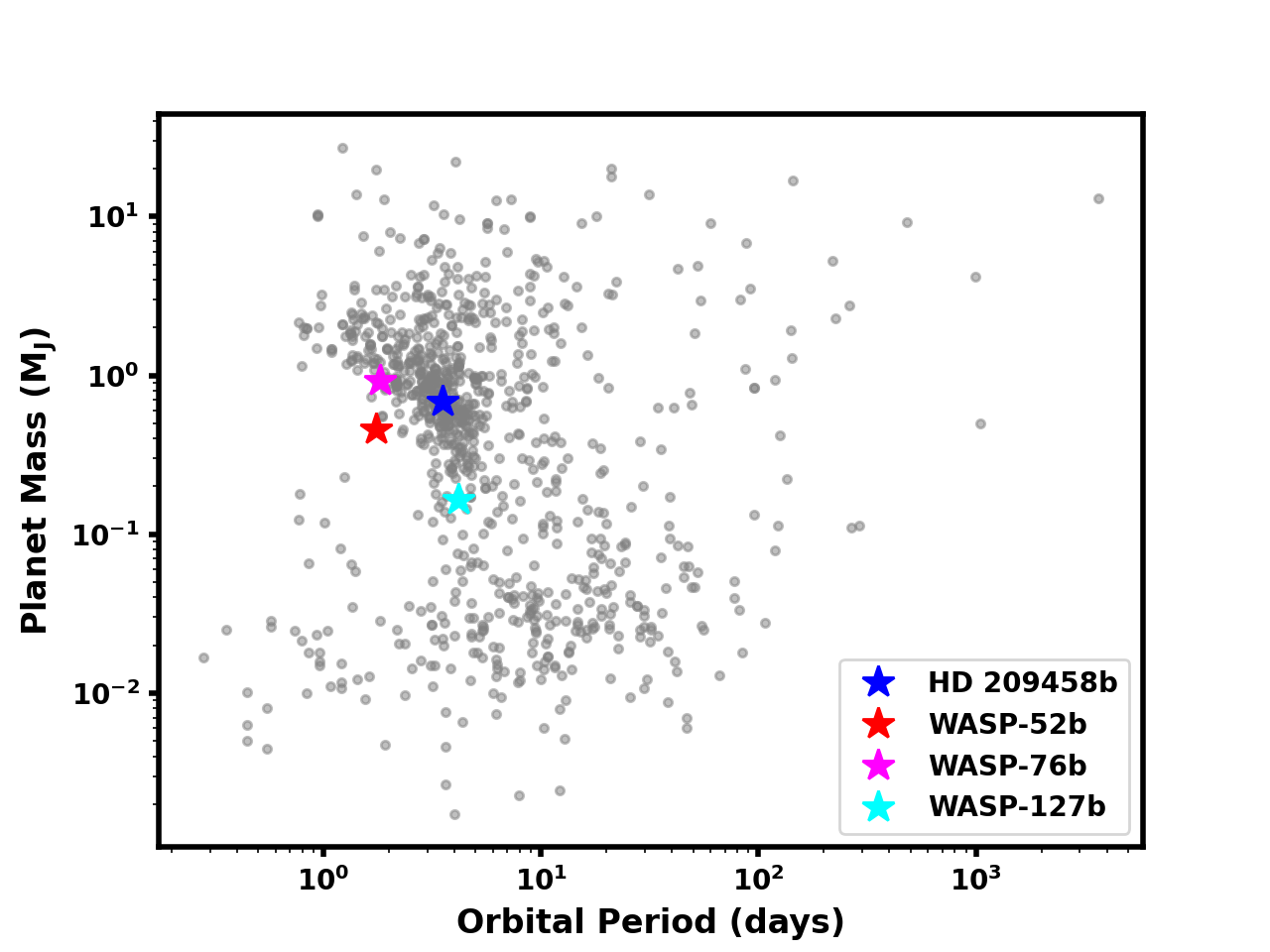} 
    \caption{Masses of known transiting exoplanets (grey dots) as a function of their orbital period. The four targets analyzed in this work are shown as colored stars, as in the figure's legend. Data obtained from the NASA Exoplanet Archive (\citealt{Akeson2013}) on July 30th 2024.}  
    \label{fig: targets}
\end{figure}
\begin{table*}
\centering 
\caption{Observing log for the ESPRESSO observations used within this work.}\label{tab: ESPRESSOdata} 
\begin{tabular}{l c c c c c c c c} \hline \hline
\noalign{\smallskip}
Planet & Date of & time  & Airmass$^{(a)}$ & $T_\mathrm{exp}$ & $N_\mathrm{obs}$ & $S/N^{(b)}$ & $S-$index $^{(c)}$ &DRS version\\
 & observation & [UT] & & [s] & (in/out) & \ion{Na}{I} order &  & \\ \hline 
 \noalign{\smallskip}
HD 209458b & 2019-07-20 & 03:43-09:30 & 2.23-1.38-1.89 & 175 & 89 (46/43) & 122-261 & 0.127-0.129& 2.0.0\\ 
 & 2019-09-11 & 00:35-06:06 & 2.00-1.38-1.93 & 175 & 85 (46/39) & 132-228 & 0.128-0.130 & 2.0.0 \\ 
WASP-52b & 2018-11-01 & 00:51-04:37 & 1.21-1.20-1.87 & 500 & 25 (12/13) & 22-30 & 0.467-0.508 & 1.3.2 \\ 
& 2018-11-08 & 00:05-04:24 & 1.22-1.20-1.98 & 800 & 19 (9/10) & 24-36 & 0.481-0.511 & 1.3.2 \\ 
& 2018-11-15 & 00:23-04:13 & 1.20-1.20-2.21 & 800 & 16 (8/8) & 26-38 & 0.484-0.504 & 1.3.2 \\ 
WASP-76b & 2018-09-03 & 03:49-09:56 & 2.15-1.13-1.37 & 600 & 34 (19/13) & 52-117 & 0.111-0.114 & 2.0.0 \\  
 & 2018-10-31 & 01:11-08:13 & 1.48-1.13-2.64 & 300 & 70 (37/33) & 64-82 & 0.110-0.114 & 2.0.0 \\ 
WASP-127b & 2019-02-25 & 02:00-08:56 & 1.53-1.07-1.95 & 300 & 69 (42/25) & 23-62 & 0.118-0.130 & 2.0.0\\ 
& 2019-03-18 & 00:02-07:28 & 1.84-1.07-1.88 & 300 & 74 (42/31) & 30-53 & 0.118-0.127 & 2.0.0\\ 
\hline 
\end{tabular}\\
\begin{flushleft}
\textbf{Notes:}\\
$^{(a)}$ First and third values refer to the airmass at the beginning and the end of the observation, respectively. The second value is the minimum airmass.\\
$^{(b)}$ Minimum and maximum S/N for each observation, calculated in the \ion{Na}{I} echelle order.\\
$^{(c)}$ Minimum and maximum \ion{Ca}{II} H\&K S-index for each observation, computed with the \texttt{ACTIN2} code (\citealt{Gomes2018, Gomes2021}). 

\end{flushleft}
\end{table*}
\begin{table*}
\centering  
\small 
\caption{Stellar, planetary, and orbital parameters adopted in this work for the four ESPRESSO targets.}\label{tab: systemparams1} 
\begin{tabular}{l c c c c} \hline \hline
\noalign{\smallskip}
Parameter & HD 209458b & WASP-52b  & WASP-76b & WASP-127b\\ \hline 
\noalign{\smallskip}
\textbf{Star} & & & &\\
\noalign{\smallskip}
$M_\star$ (M$_\odot$) & $1.119 \pm 0.033^{(a)}$ & $0.87 \pm 0.03^{(g)}$ & $1.458 \pm 0.021^{(i)}$ & $1.31 \pm 0.05^{(k)}$\\ 
$R_\star$ (R$_\odot$) & $1.155 \pm 0.016^{(a)}$ & $0.79 \pm 0.02^{(g)}$ & $1.756 \pm 0.071^{(i)}$ & $1.33 \pm 0.03^{(k)}$ \\ 
$K_\star$ (m s$^{-1}$) & $84.27^{+0.69}_{-0.70}$ $^{(b)}$ & $82.8 \pm 2.1^{(h)}$ & $116.02^{+1.29}_{-1.35}$ $^{(i)}$ & $21.51 \pm 2.78^{(l)}$\\ 
$T_\mathrm{eff}$ (K) & $5996^{+20}_{-10}$ & $5051^{+23}_{-29}$ & $6220 ^{+47}_{-53}$ & $5704^{+32}_{-11}$ \\
$\log g$ (cm s$^{-2}$) & $4.237 \pm 0.003$ & $4.525^{+0.010}_{-0.007}$ & $4.007 ^{+0.028}_{-0.037}$ & $4.230^{+0.013}_{-0.003}$ \\
$\mathrm{[Fe/H]}$ (dex) & $0.0 \pm 0.2$ & $0.28 \pm 0.02$ & $0.3 \pm 0.2$ & $-0.38 \pm 0.05$ \\ 
$A(\mathrm{Na})$ (dex) & $6.21 \pm 0.03$ & $6.41 \pm 0.03$ & $6.79 \pm 0.02$ & $5.77 \pm 0.06$ \\
$\mathrm{[Na/Fe]}$ (dex) & $0.16 \pm 0.03$ & $-0.12 \pm 0.03$ & $0.38 \pm 0.02$ & $0.00 \pm 0.06$ \\ 
$v \ \sin i_\star$ (km s$^{-1}$) & $4.228 \pm 0.007^{(b)}$ & $2.622 \pm 0.074^{(h)}$ & $1.48 \pm 0.28$ $^{(i)}$ & $0.53^{+0.07}_{-0.05}$ $ ^{(m)}$\\
\noalign{\smallskip}
\textbf{Planet} & & & & \\
\noalign{\smallskip}
$M_\mathrm{p}$ (M$_\mathrm{J}$) & $0.685 \pm 0.015^{(a)}$ & $0.46 \pm 0.02^{(g)}$ & $0.894^{+0.014}_{-0.013}$ $^{(i)}$ & $0.165 \pm 0.021^{(l)}$\\
$R_\mathrm{p}$ (R$_\mathrm{J}$) & $1.359 \pm 0.019^{(a)}$ & $1.27 \pm 0.03^{(g)}$ & $1.854 ^{+0.077}_{-0.076}$ $^{(i)}$ & $1.311 \pm 0.025^{(l)}$\\
$K_\mathrm{p}$ (km s$^{-1}$) & $145.0 \pm 1.5$ & $168.6 \pm 1.9$ & $198.4 \pm 1.2$ & $125.9 \pm 3.5$ \\ 
$i_\mathrm{p}$ (deg) & $86.78 \pm 0.07^{(c)}$ & $85.35 \pm 0.20^{(g)}$ & $89.623^{+0.005}_{-0.034}$ $^{(i)}$ & $87.85 \pm 0.35^{(m)}$ \\
$\lambda$ (deg) & $1.58 \pm 0.08^{(b)}$ & $1.1 \pm 1.1^{(h)}$ & $61.28 ^{+7.61}_{-5.06}$ $^{(i)}$ & $-128.41 ^{+5.60}_{-5.46}$ $^{(m)}$ \\
\noalign{\smallskip}
\textbf{System} & & & & \\
\noalign{\smallskip}
$P_\mathrm{orb}$ (days) & $3.52474859 \pm 0.00000038^{(d)}$ & $1.7497798 \pm 0.0000012^{(g)}$ & $1.80988198^{+0.00000064}_{-0.00000056}$ $^{(i)}$ & $4.17806203 \pm 0.00000088^{(l)}$ \\
$T_\mathrm{c}$ (BJD) & $2454560.80588 \pm 0.00008^{(c)}$ & $2455793.68143 \pm 0.00009^{(g)}$ & $2458080.626165^{+0.000418}_{-0.000367}$ $^{(i)}$ & $2 456 776.621238 \pm 0.00023181^{(l)}$\\
$T_{14}$ (days) & $0.124 \pm 0.002^{(e)}$ & $0.0754 \pm 0.0005^{(g)}$ & $0.1564 \pm 0.0005 ^{(j)}$ & $0.18137185 \pm 0.00035381^{(l)}$\\
$a$ (AU) & $0.04707 \pm 0.00047^{(a)}$ & $0.0272 \pm 0.0003^{(g)}$ & $0.0330 \pm 0.0002^{(i)}$ & $0.04840^{+0.00136}_{-0.00095}$ $^{(l)}$\\
$v_\mathrm{sys}$ (km s$^{-1}$) & $-14.741 \pm 0.002^{(f)}$ & $-0.8212 \pm 0.0007^{(h)}$ & $-1.11 \pm 0.50^{(i)}$ & $-9.29546 \pm 0.0014^{(m)}$\\
 & & $-0.8434 \pm 0.0006^{(h)}$ & & \\
 & & $-0.8357 \pm 0.0006^{(h)}$ & & \\
 \noalign{\smallskip}
\hline 
\end{tabular}\\
\begin{flushleft}
\textbf{Notes:}\\
$M_\star$, $R_\star$ and $M_\mathrm{p}$, $R_\mathrm{p}$ are the mass and radius of the star and planet, respectively. $K_\star$ is the semi-amplitude of the stellar RV signal, derived from ESPRESSO data. 
All values for the stellar effective temperature and surface gravity ($T_\mathrm{eff}$, $\log{g}$) are taken from Gaia DR3 catalogue. 
The metallicity values ([Fe/H]) are sourced from Gaia DR3, but the uncertainties have been adjusted to account for the level of agreement or disagreement between the Gaia BP/RP and RVS values and those reported in the literature. 
Eccentricity is fixed to 0 for all the systems. The two entries for the sodium abundance,  $A(\mathrm{Na})$ and $\mathrm{[Na/Fe]}$, are inferred by fitting the 3D NLTE synthetic spectra to the Na D lines and the 6154/6160\,\r{A} lines, respectively (see Sect. \ref{sec: synthspectra}).
$v \ \sin i_\star$ is the stellar rotational velocity. $K_\mathrm{p}$ is the planet RV semi-amplitude (see Sect. \ref{sec: transspectr}). The inclination and the projected obliquity of the orbit are $i_\mathrm{p}$ and $\lambda$, respectively. The other parameters represent: the orbital period ($P_\mathrm{orb}$), the time of conjunction ($T_\mathrm{c}$), the transit duration ($T_{14}$), the semi-major axis ($a$) and the systemic velocity ($v_\mathrm{sys}$).
$^{(a)}$ \citet{Torres2008}; $^{(b)}$ \citet{Casasayas2021}; $^{(c)}$ \citet{Evans2015}; $^{(d)}$ \citet{Knutson2007}; $^{(e)}$ \citet{Richardson2006}; $^{(f)}$ \citet{Naef2004}; $^{(g)}$ \citet{Hebrard2013}; $^{(h)}$ \citet{Chen2020}; $^{(i)}$ \citet{Ehrenreich2020}; $^{(j)}$ \citet{Fu2021}; $^{(k)}$ \citet{Lam2017}; $^{(l)}$ \citet{Seidel2020a}; $^{(m)}$ \citet{Allart2020}
\end{flushleft}
\end{table*}

For this work, we employed observations from the high-resolution ground-based spectrograph ESPRESSO at the Very Large Telescope (VLT; \citealt{Pepe2021}), with a spectral resolving power of $R \approx 140 \ 000$, and covering the wavelength range $380-788$\,nm.  
The data are summarized in Table \ref{tab: ESPRESSOdata}, where the observing logs are reported. Further information about discarded datasets that would adversely affect the transmission spectra, due to low signal-to-noise ratio (S/N) or bad weather conditions for instance, can be found in Sect. \ref{sec: HD209intro} to \ref{sec: WASP127intro}. The four analyzed targets are shown in Fig. \ref{fig: targets}.  
The ESPRESSO Data Reduction Software (DRS) pipeline, as reported in Table \ref{tab: ESPRESSOdata}, was used to reduce all the observations and extract the one-dimensional spectra used for the analysis. 

The stellar, planetary, and orbital parameters of the four star-planet systems are shown in Table \ref{tab: systemparams1}.

\subsection{HD 209458b}\label{sec: HD209intro}
HD 209458b is a benchmark hot Jupiter orbiting around a bright G-type star. It is the first planet found to be transiting its host star (\citealt{Charbonneau2000}; \citealt{Henry2000}) and one of the most extensively studied in the literature to date. \ion{Na}{I} absorption in transmission spectra has been highly debated by comparing observations at high- and low-resolution with different instruments and telescopes (e.g., \citealt{Morello2022}). 
  
After the first detection in the early 2000s from low-resolution observations taken with the Space Telescope Imaging Spectrograph (STIS) on board the HST (\citealt{Charbonneau2002}), upper limits to the \ion{Na}{I} absorption were set from the ground a few years later (\citealt{Snellen2004}; \citealt{Narita2005}). Over the years, numerous other detections have been reported from both ground- and space-based observations (e.g., \citealt{Sing2008}; \citealt{Snellen2008}; \citealt{Albrecht2009}; \citealt{LanglandShula2009}; \citealt{Jensen2011}; \citealt{Astudillo2013}; \citealt{Santos2020}), confirming the presence of \ion{Na}{I} at both low and high resolution. However, in 2020, the first doubts emerged when \citet{Casasayas2020} analyzed high-resolution data from the CARMENES and HARPS-N spectrographs and suggested that the spectral features observed in the transmission spectrum of the \ion{Na}{I} D lines might actually be due to the combined effects of the CLV and RM effect of the star, rather than the absorption from the planet. This non-detection was then supported by two nights of high-resolution, high-S/N observations with ESPRESSO (\citealt{Casasayas2021}; \citealt{Morello2022}; \citealt{Dethier2023}). Indeed, in the analysis of these data, \citet{Casasayas2021} argued that the spectral features in transmission spectra at 589\,nm could be explained by the CLV and RM effects of the star. However, the modeled stellar spectra were unable to perfectly reproduce the observed spectral feature, especially in the transmission LCs of the \ion{Na}{I} D lines.   
They indeed emphasize the necessity of utilizing precise stellar models to correctly account for the CLV and RM effect, which might otherwise bias the analysis of transit spectroscopy observations.

In this work, we reanalyze two transits observed with ESPRESSO under the ESO program 1102.C-0744 in July and September 2019.
The datasets are the same that have been previously analyzed in \citet{Casasayas2021} (CB21 from hereafter). 
For this target, all exposures listed in Table \ref{tab: ESPRESSOdata} were used, and none were discarded.

\subsection{WASP-52b}\label{sec: WASP52intro}
WASP-52b is a highly inflated hot Jupiter around a K-type, faint, moderately active, star (see the S-index in Table \ref{tab: ESPRESSOdata})
with a strong CLV+RM effect (e.g., \citealt{Chen2020}) and with occulted starspots and faculae observed in transit LCs (e.g., \citealt{Kirk2016}; \citealt{Mancini2017}). \ion{Na}{I} has been detected by \citet{Chen2020} by combining three transits observed at high resolution with ESPRESSO in October and November 2018 under the ESO programs 0102.D-0789 and 0102.C-0493. 
We reanalyzed those data. From the first two nights, we discarded the last exposure due to an extremely low signal-to-noise ratio ($\mathrm{S/N < 2}$).

Several activity indicators were used to monitor WASP-52 during the transit observations, but no excess absorption suggesting activity was registered (see Sect. 5.1 in \citealt{Chen2020}). 
Also, \citet{Cegla2023} did not detect any evidence of spot occultations during the same transits using simultaneous photometric observations from EulerCam (\citealt{Lendl2012}) and the Next Generation Transit Survey (NGTS; \citealt{Wheatley2018}).

The atmosphere of WASP-52b is the first one analyzed with the ESPRESSO spectrograph. 
\ion{Na}{I} was previously detected from low-resolution observations from HST (e.g., \citealt{Alam2018}) and ground-based telescopes spanning from the optical to the infrared wavelengths (e.g., \citealt{Chen2017}; \citealt{Louden2017}; \citealt{Bruno2018}), as well as with combined data from different instruments (e.g., \citealt{Bruno2020}). Most of these studies found evidence for a cloudy atmosphere, and specifically an optically thick and gray cloud layer at high altitudes.

\subsection{WASP-76b}\label{sec: WASP76intro}
WASP-76b is an exoplanet belonging to the class of ultra-hot Jupiters (UHJs), which are highly irradiated gas giants with a surface temperature greater than 2000 K and an atmosphere rich in atomic species. It orbits an F-type star with an orbital period of only 1.8 days. Different atmospheric chemistry is expected between the dayside and the nightside of this kind of exoplanet, and indeed asymmetric absorption signatures have been detected during transits of WASP-76b by \citet{Ehrenreich2020}. The latter measured a blueshifted \ion{Fe}{I} line of about $-11.0 \pm 0.7$\,km\,s$^{-1}$, attributed to a combination of winds and planetary rotation. The data of the two transits observed by ESPRESSO in 2018, and reanalyzed in this work, were part of the ESO program 1102.C-744. In the analysis, we discarded the last exposure of the second transit due to a very low S/N (S/N $\approx 20$). 
These observations were also analyzed in \citet{Tabernero2021}, where they found a strong detection of \ion{Na}{I} absorption in the transmission spectra, but did not correct for the CLV+RM effect since they estimated the amplitude of the effects to be much smaller than the error bars on the datapoints. \ion{Na}{I} was also previously detected in HARPS data (e.g., \citealt{Seidel2019}; \citealt{Langeveld2022}) as a broadened feature of the \ion{Na}{I} doublet but with a lower significance than the ESPRESSO data. Then, combining HARPS and ESPRESSO data, \citet{Seidel2021} could distinguish between different wind patterns in the lower and upper atmospheres, as well as retrieve the temperature profile of WASP-76b by analyzing the broadened line shape of the \ion{Na}{I} doublet. 
Finally, the same data were analyzed by \citet{Kesseli2022} using the cross-correlation technique, which revealed asymmetries in radial velocities (RVs) between ingress and egress in most of the detected species, including \ion{Na}{I}, and provided insights into the atmospheric dynamics of this planet.

\subsection{WASP-127b}\label{sec: WASP127intro} 
WASP-127b is a close-in, highly inflated super-Neptune in a misaligned retrograde orbit around a bright, slowly rotating G-type star. Low-resolution, ground-based observations first revealed a very peculiar atmosphere, rich in alkali metals (\ion{Na}{I}, \ion{Li}{I} and \ion{K}{I}) but also with unexpected signatures of water (\citealt{Palle2017}; \citealt{Chen2018}). 
After that, the presence of \ion{Na}{I} was debated in \citet{Seidel2020a}, who detected the D$_2$ line but not the D$_1$ line in high-resolution HARPS data. However, \ion{Na}{I} absorption was again detected at low resolution in a recent analysis of combined HST and Spitzer Space Telescope data (\citealt{Spake2021}). 
Moreover, due to its peculiar properties, this exoplanet has been extensively studied since its discovery and was also recently observed by the James Webb Space Telescope (JWST; \citealt{Barstow2015}) in May and December 2023, under the programs GO 2437 and GTO 1201, respectively.
ESPRESSO observed two transits of this exoplanet in February and March 2019 under the ESO program 1102.C-0744, and the data were analyzed in \citet{Allart2020}. In their work, they did not correct the transmission spectrum for the CLV+RM effect since the model showed an extremely small amplitude due to the unusual planetary orbit. Indeed, they were still able to retrieve a \ion{Na}{I} absorption at 9$\sigma$ confidence level, blueshifted of $2.74 \pm 0.79$\,km\,s$^{-1}$, interpreted as a signal of the dynamics of the planet atmosphere, specifically as winds moving from the day- to the night-side.
We reanalyzed those ESPRESSO data without discarding any exposure.

\section{Synthetic stellar spectra}\label{sec: synthspectra}

\begin{figure}
    \centering
    \includegraphics[width=9cm]{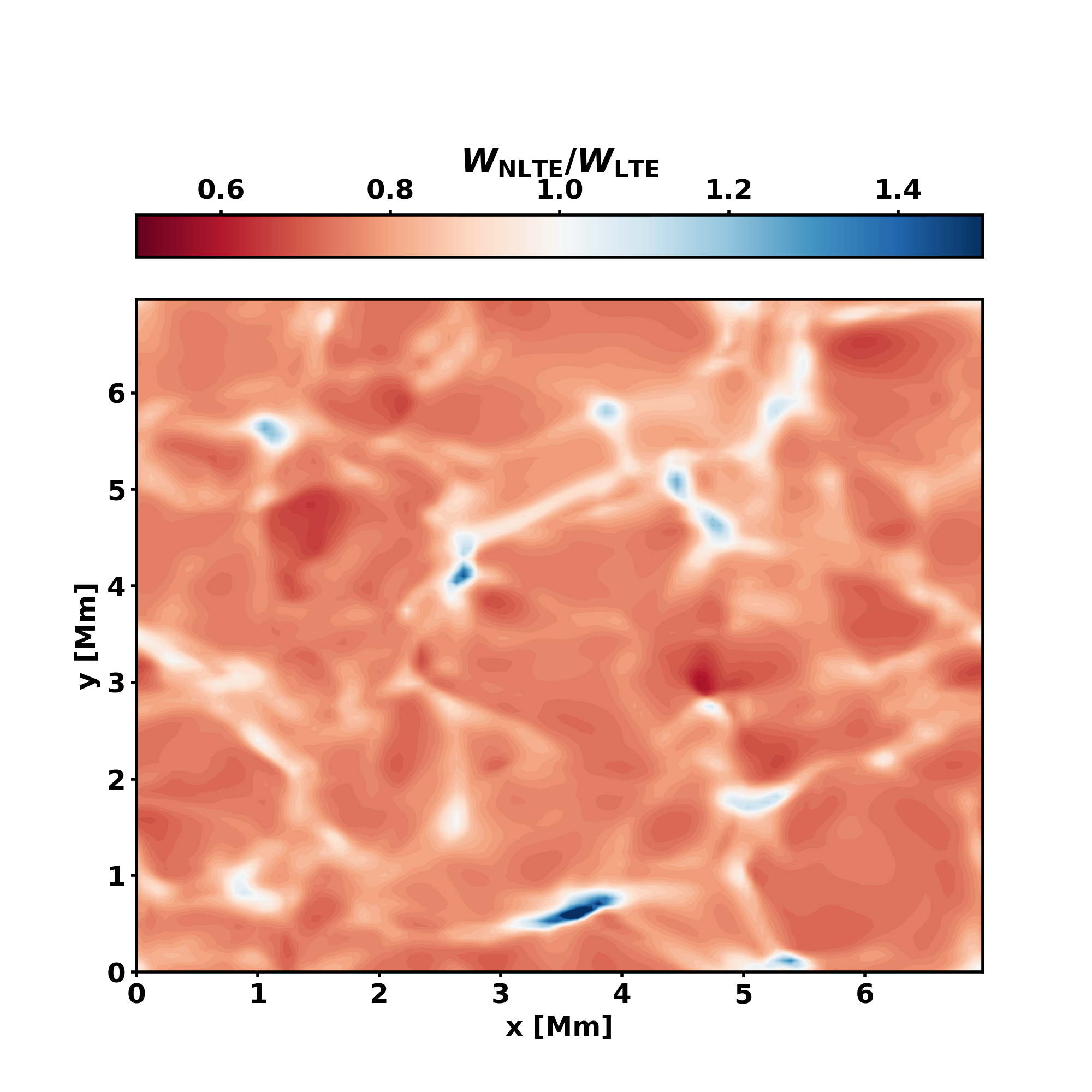} 
    \caption{Ratio of the NLTE to LTE equivalent width of \ion{Na}{I} 5896\,\r{A} for a single \textsc{Stagger} snapshot of a star with $T_\mathrm{eff}=6000$\,K, $\log g=4.5$\,dex, and [Fe/H]=0.0. In the up-flowing granules, overionization makes the line weaker in NLTE.}  
    \label{fig: stagger}
\end{figure}
\begin{figure}
    \centering
    \includegraphics[width=1.0\linewidth]{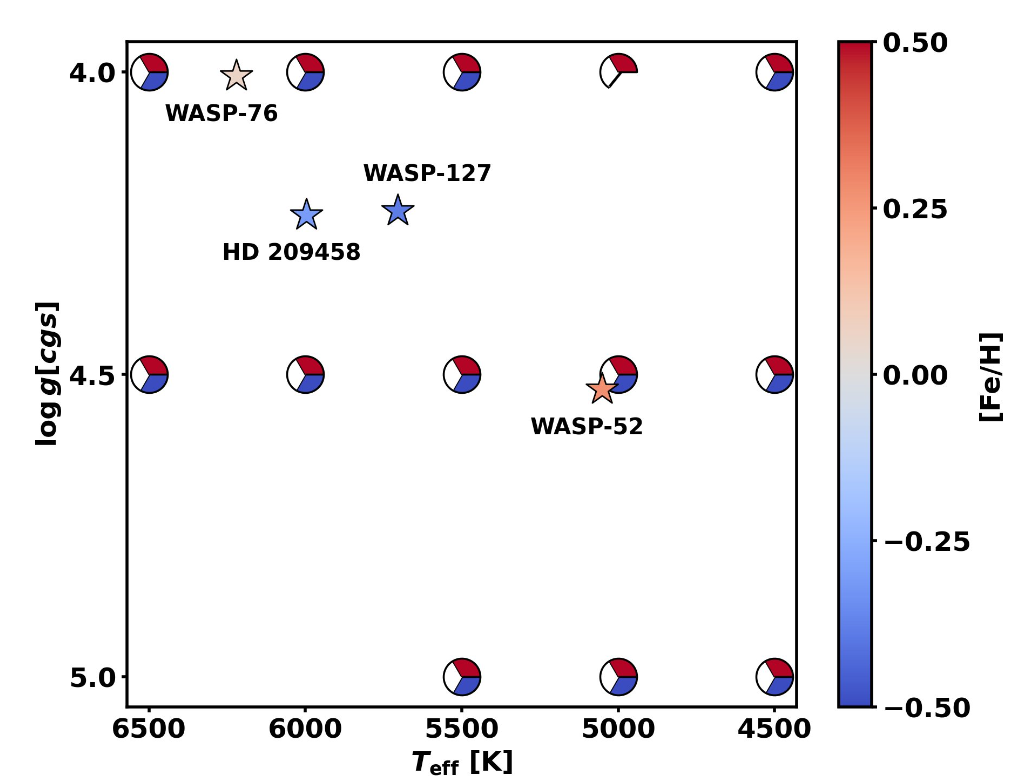}
    \caption{HR diagram illustrating the \textsc{Stagger}-grid nodes at which the 3D NLTE computations have been performed, for [Fe/H]$= -0.5$ (blue), 0.0 (white), and $+0.5$ (red). Additionally, the four host stars of the gas giants analyzed in this work are overplotted, colored by their metallicity.}
    \label{fig: HRdiagram}
\end{figure}
   \begin{figure*}
   \resizebox{\hsize}{!}
            {\includegraphics[width=10cm]{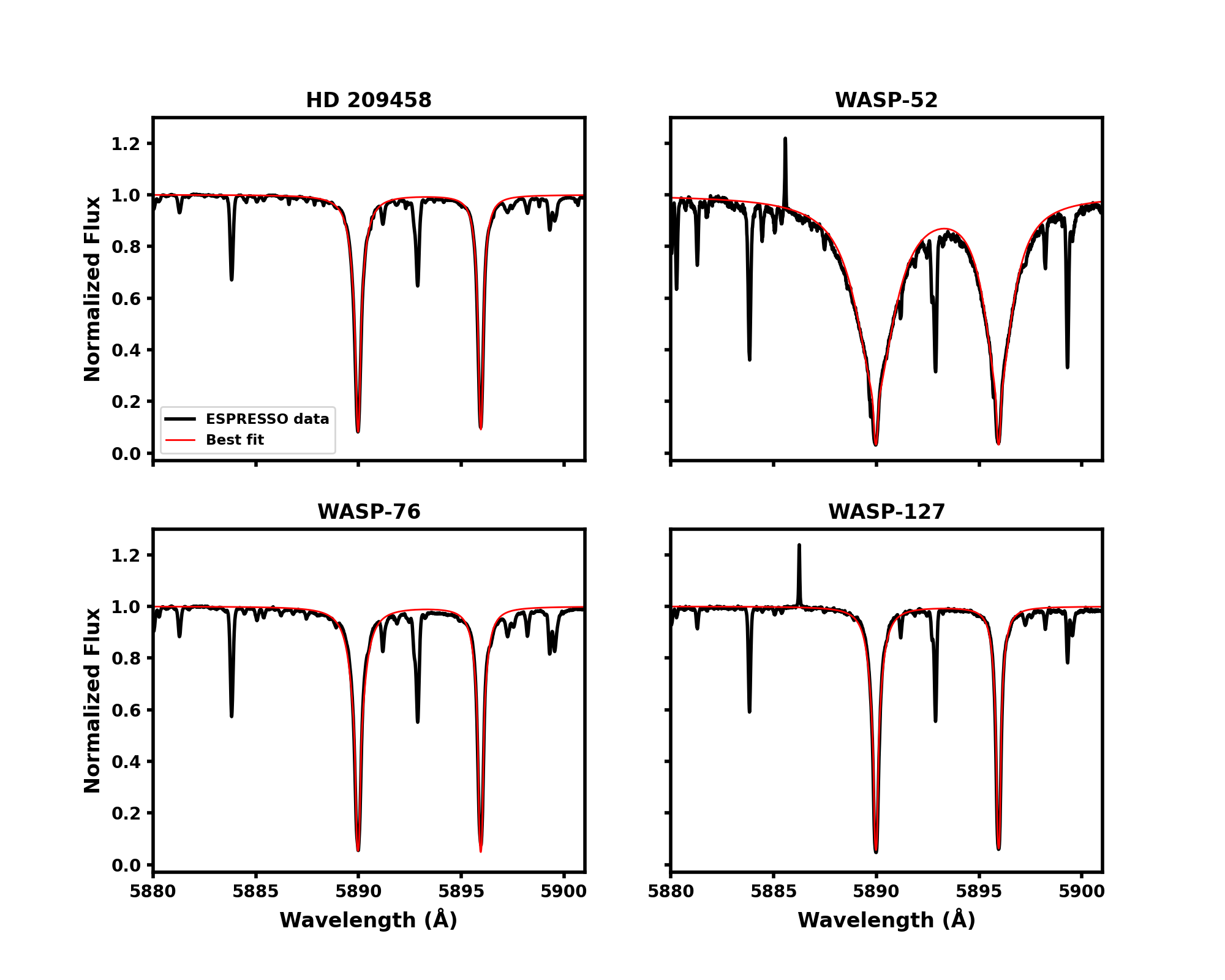}}
            
      \caption{Master Out (in black) around the \ion{Na}{I} doublet region with overplotted the best-fit synthetic 3D NLTE spectra (in red) for: HD 209458 (upper left panel), WASP-52 (upper right panel), WASP-76 (middle left panel), WASP-127 (middle right panel). The two emission spikes in WASP-52 and WASP-127 are due to cosmic rays.} 
              
         \label{fig: FoutNa}
   \end{figure*}
The synthetic stellar spectra used in the calculations for the CLV+RM model (see Sect. \ref{sec: CLVRM}), have been computed with the radiative transfer MPI-parallelised code \texttt{Balder} (\citealt{Amarsi2018}). The latter originates from the \texttt{Multi3d} code (\citealt{Botnen1999}; \citealt{Leenarts2009}), which is able to solve the restricted NLTE problem for trace elements, for user-specified atomic elements, and for model atmospheres in either 1D or 3D. \texttt{Balder} can also synthesize spectra under the assumption of LTE line formation.
The equation-of-state, the background line opacities, and the continuous opacities are calculated using the \texttt{BLUE} package, as described in \citet{Amarsi2016, Amarsi2016b}.  
\texttt{Balder} solves the radiative transfer equation and the statistical equilibrium equations simultaneously in an iterative way, until convergence of the level populations is reached. 
When performing radiative transfer in 3D, the computation is performed along several short characteristic rays at different $\mu$ inclinations and azimuthal $\phi$ angles. In this work, the following number of rays was used with the Lobatto quadrature: $n_\mu=8$ and $n_\phi=4$. 
The details about the model atmospheres and the model atom used are given in Sect. \ref{sec: modelatmos} and \ref{sec: modelatom}, respectively.
The computation of the synthetic spectra is described in Sect. \ref{sec: synthspectra2}.

\subsection{Model atmosphere}\label{sec: modelatmos}
Snapshots from 3D radiation-hydrodynamics (RHD) simulations calculated with the \textsc{Stagger}-code (\citealt{Galsgaard1995}; \citealt{Stein1998}; \citealt{Collet2011}; \citealt{Magic2013}; \citealt{Collet2018}; \citealt{Chiavassa2018}; \citealt{Stein2024}) are employed as model atmospheres for the spectrum synthesis performed in \texttt{Balder}. The detailed 3D NLTE radiative transfer was calculated on five snapshots of each model from the recently updated \textsc{Stagger}-grid (\citealt{RodriguezDiaz2023}) adequately spaced so as to capture several convective turnovers. 
The snapshots were selected according to the procedure described in Sect. 4.4 of \citet{RodriguezDiaz2023}, ensuring that the average hydrodynamic gas properties remain consistent with the full time series.
The resulting spectra (intensities at different $\mu$-angles) from the five chosen snapshots were then temporally and spatially averaged. We refer the reader to \citet{RodriguezDiaz2023} for more details about the code and the new \textsc{Stagger}-grid.

The 3D RHD \textsc{Stagger}-model is set on a cartesian grid of size 240$\times$240$\times$240, corresponding to a physical size that varies from star to star, but chosen in such a way to always include at least ten granules in the simulation. 
However, in order to decrease the computational time for the spectral synthesis, all the atmospheric snapshots were resized to a resolution of 48$\times$48$\times$240. The vertical dimension indeed, is not equidistant in space but a higher resolution is adopted in the line-forming region (i.e., photosphere). \citet{RodriguezDiaz2023} showed that reducing the resolution of the horizontal dimension does not significantly affect the resulting spectral line.  

Figure \ref{fig: stagger} displays an example of the spatially-resolved 3D NLTE to 3D LTE ratio of the equivalent width ($W_\lambda$) of the \ion{Na}{I} 5896\,\r{A} line at the surface of one of the \textsc{Stagger} snapshots used in this work. NLTE effects are strongly line-strength dependent, and since $W_\lambda$ represents the strength of a spectral line independently of the spectral resolution, it is a good indicator to highlight discrepancies between LTE and NLTE modeling. 
From the figure, it can be noticed how the NLTE effects vary over the convection pattern: overpopulation (i.e., $W_{NLTE}/ W_{LTE} > 1$) in the intergranular lanes, and under-population in the granules. 
Indeed, the steeper temperature gradient of the granules leads to a larger split between the mean radiation field $J_\nu$ and the Planck function $B_\nu$, as well as more efficient overionization. This effect was observed in the Sun for the \ion{Na}{I} (\citealt{Canocchi2024}) as well as for other elements (e.g., \citealt{Asplund2004}; \citealt{Lind2017}; Mallinson et al. in prep.). 

In 3D atmospheres, granulation on the stellar surface arises naturally: the granules are the locations of upflowing hot plasma and gas, whereas the dark intergranular lanes are the locations of downflowing cooler plasma. Analogously, effects such as spectral line broadening and line asymmetries are naturally accounted for in the model itself (e.g., \citealt{Dravins1981}; \citealt{Asplund2000}; \citealt{Dravins2021}) without the need for tunable parameters such as micro- and macroturbulence ($v_\mathrm{mic}$ and $v_\mathrm{mac}$) typically used in 1D stationary, hydrostatic models. 
For comparison and completeness, also synthetic spectra from 1D \texttt{MARCS} (\citealt{Gustafsson2008}) model atmospheres have been computed and used throughout this work, adopting $v_\mathrm{mic}=1.0$\,km s$^{-1}$ and $v_\mathrm{mac}=3.5$\,km s$^{-1}$, respectively.

\subsection{Model atom}\label{sec: modelatom}
The strong NLTE effects of the \ion{Na}{I} atom have been reported in several studies in which NLTE spectra were necessary to reproduce observed data of the Sun and other stars (e.g., \citealt{Mashonkina2000}; \citealt{Lind2011}; \citealt{Asplund2021}; \citealt{Canocchi2024}). Specifically, the main NLTE effect of the resonance \ion{Na}{I}\,D lines in 1D is that they are affected by over-recombination of the ground state and a sub-thermal source function, resulting in a stronger line in NLTE compared to LTE at the same abundance.  
For more details about the NLTE effects of \ion{Na}{I}, we refer the reader to Sect. 3.1 of \citet{Lind2011} and Sect. 3.2 of \citet{Canocchi2024}.

The model atom used in this work as input for the \texttt{Balder} NLTE calculations, is the one developed and tested in \citet{Lind2011}, with modifications as described in \citet{Canocchi2024}. 
It consists of a total of 23 levels: 22 energy levels for \ion{Na}{I} and one for the continuum (\ion{Na}{II}). The model includes both radiative and collisional transitions among these energy levels, with detailed quantum mechanical calculations specifically for inelastic collisions involving Na and hydrogen atoms (\citealt{Belyaev2010}; \citealt{Barklem2010}) as well as electrons (\citealt{Park1971}; \citealt{Allen1993}; \citealt{Igenbergs2008}).
The \ion{Na}{I} doublet lines arise from transitions between the 3s--3p$_{1/2}$ and 3s--3p$_{3/2}$ levels.

\subsection{Synthetic spectra}\label{sec: synthspectra2}
For a given model atom and model atmosphere, radiative transfer calculations were performed independently for three different values of sodium abundances ($A(\mathrm{Na})$), defined as:
\begin{equation}
    A(\mathrm{Na})= \mathrm{log} \epsilon_\mathrm{Na} = \mathrm{log}_{10} (N_\mathrm{Na}/N_\mathrm{H}) + 12 \ ,
    \label{eq: Ab}
\end{equation}
where $N_\mathrm{Na}$ and $N_\mathrm{H}$ are the number densities of Na and H.  
In particular, for each stellar metallicity (i.e., $\mathrm{[Fe/H]= \log \epsilon_\mathrm{Fe}-\log \epsilon_\mathrm{Fe_\odot}}$), we synthesize spectra for the following sodium abundances: [Na/Fe]\footnote{[Na/Fe]=[Na/H]-[Fe/H]}= -0.5, 0.0, and 0.5, that is within the range of observed sodium abundance in large spectroscopic stellar surveys such as GALAH (e.g., \citealt{GALAHSurvey}; \citealt{Buder2021}). 
The synthetic spectra have a resolving power of $R \approx 500 \ 000$ in the \ion{Na}{I} D region, covering the wavelength range 5866.0--5920.0\,\r{A}. 
Therefore, a grid of 3D NLTE synthetic spectra has been computed, for the following stellar parameters: $T_\mathrm{eff}$ between 4500 and 6500\,K in steps of 500\,K; $\log g$= 4.0, 4.5, and 5.0 dex; [Fe/H]=-0.5, 0.0, and +0.5 dex. These parameters cover most of the FGK-type stars hosting exoplanets observed in transmission spectroscopy studies.
Figure \ref{fig: HRdiagram} shows the Hertzsprung-Russell (HR) diagram with the extent of the computed stellar grid and the locations of the host stars analyzed in this work. In addition, the stellar parameters of the synthetic spectra computed in this work are reported in Table \ref{tab: synthspectraparams} in Appendix.
Seven models are not available in the \textsc{Stagger}-grid: $T_\mathrm{eff}=6000$ and 6500\,K with $\log g=5.0$ for all three metallicities, and $T_\mathrm{eff}=5000$\,K, $\log g=4.0$, $\mathrm{[Fe/H]}=-0.5$.

We make the grid of synthetic spectra publicly available, together with a python interpolation routine to get the flux and/or intensity spectra at different $\mu$-angles for given stellar parameters of $T_\mathrm{eff}$, $\log g$, [Fe/H], and $A$(Na).

For each of the four targets in this work, we use a simple trilinear interpolation to interpolate the flux synthetic spectra from the grid to the specific $T_\mathrm{eff}$, $\log g$, and [Fe/H] of the host star, using the stellar parameters reported in Table \ref{tab: systemparams1}. 
The \textsc{Stagger}-grid, as shown in Fig. \ref{fig: HRdiagram}, is sparse compared to standard 1D grids like \texttt{MARCS} (\citealt{Gustafsson2008}), and some of our target stars are located quite far from the grid nodes. In a previous study on a 3D NLTE grid of synthetic spectra for \ion{Li}{I}, \citet{Wang2021} investigated the accuracy of different interpolation methods on the \textsc{Stagger}-grid, concluding that the commonly used spline interpolation performs well for spectral line profiles (see Fig. 10 in \citealt{Wang2021}). However, they suggested that more complex methods, such as those involving neural networks, should be employed for abundance determinations.
In this work, we decided to use linear interpolation because we found that spline interpolation performs well for relatively weak spectral lines such as the \ion{Li}{I} line analyzed by \citet{Wang2021}, but it overestimates the flux in the wings of very broad lines such as the \ion{Na}{I} D lines.
We plan to explore the impact of different and more accurate interpolation methods on our \ion{Na}{I} grid in a forthcoming paper, which will also extend the grid of 3D NLTE synthetic spectra to include lower metallicity models as well as giant stars.

After interpolating the synthetic fluxes from the grid to match the parameters of the four host stars, we normalize the observed out-of-transit fluxes (i.e., the Master Out, see Sect. \ref{sec: masterout}) by fitting a first-order polynomial to the local continuum in the region of the \ion{Na}{I} D lines and then dividing the data by this polynomial.
After that, the synthetic normalized flux of the \ion{Na}{I} D lines is fitted to the observed normalized Master Out via a standard $\chi^{2}$ minimization routine using the \ion{Na}{I} abundance ($A(\mathrm{Na})$, see Table \ref{tab: synthspectraparams}) as a free parameter. Linear interpolation between models is applied. 
The best-fit value of $A(\mathrm{Na})$, as inferred from the \ion{Na}{I} D lines, is reported in Table \ref{tab: systemparams1}. 
The resulting best-fit line profile is shown in Fig. \ref{fig: FoutNa}. The best-fit abundance value was then employed to compute the intensity spectra at the different $\mu$-angles used in the analysis of the transmission spectra and LCs for the computation of the CLV+RM model (see Sect. \ref{sec: CLVRM}).

Similarly, we fit the 3D NLTE synthetic flux to the \ion{Na}{I} doublet at 6154\,\r{A} and 6160\,\r{A}, which is typically used to determine sodium abundance in metal-rich stars.
The synthetic spectra have a resolving power of $R \approx 400 \ 000$ in the 6154/6160\,\r{A} region, covering the wavelength range 6148--6168\,\r{A}.
The resulting [Na/Fe] values are presented in Table \ref{tab: systemparams1}. These values are in good agreement with those inferred from fitting the \ion{Na}{I} D lines, with discrepancies ranging from 0.02\,dex for WASP-76b to 0.06\,dex for WASP-127b, which could be due to unresolved blends, slightly different continuum placement or/and various errors in atomic data.
We did not compute the CLV+RM model for these lines because they are significantly weaker than the \ion{Na}{I} D lines. As a result, the transmission spectra and LCs exhibit variations on the same order of magnitude as the continuum (e.g., 0.05\% in HD 209458), making it impossible to detect any measurable absorption depth in the planet's atmosphere using these lines.  

The stellar parameters of the four stars analyzed in this work are taken from the Gaia DR3 archive (\citealt{GaiaDR32023})\footnote{\url{https://gea.esac.esa.int/archive/}} and are mostly consistent with other values reported in the literature. 
We noted a discrepancy between the $\mathrm{[Fe/H]}$ values for HD 209458 and WASP-76 from Gaia DR3 and from high-resolution spectroscopic studies in literature. However, this discrepancy does not significantly affect our results, meaning that the [Fe/H] parameter does not play a critical role in the analysis of the CLV+RM model.

Moreover, we used the 3D NLTE H$\alpha$ line profiles from the public grid of \citet{Amarsi2018} as diagnostics to validate the effective temperature. 
The extremely good agreement of the wings of the \ion{Na}{I} D lines with these models is shown in Fig. \ref{fig: Halpha} in Appendix, where the models are overplotted onto the out-of-transit spectra of the four targets. 

In Sect. \ref{sec: analysis}, we apply the 3D NLTE synthetic spectra to the high-resolution spectroscopic observations of four bright exoplanet host stars, presented in Sect. \ref{sec: obs}. We make a comparison with synthetic spectra computed with \texttt{Balder} in 3D LTE, 1D LTE, and 1D NLTE as well, and we test the improvement, if any, in the (non-)detectability of \ion{Na}{I} in their atmospheres.

\section{Search for Na absorption}\label{sec: methods} 
\begin{figure*}
   \resizebox{\hsize}{!}
            {\includegraphics[width=10cm]{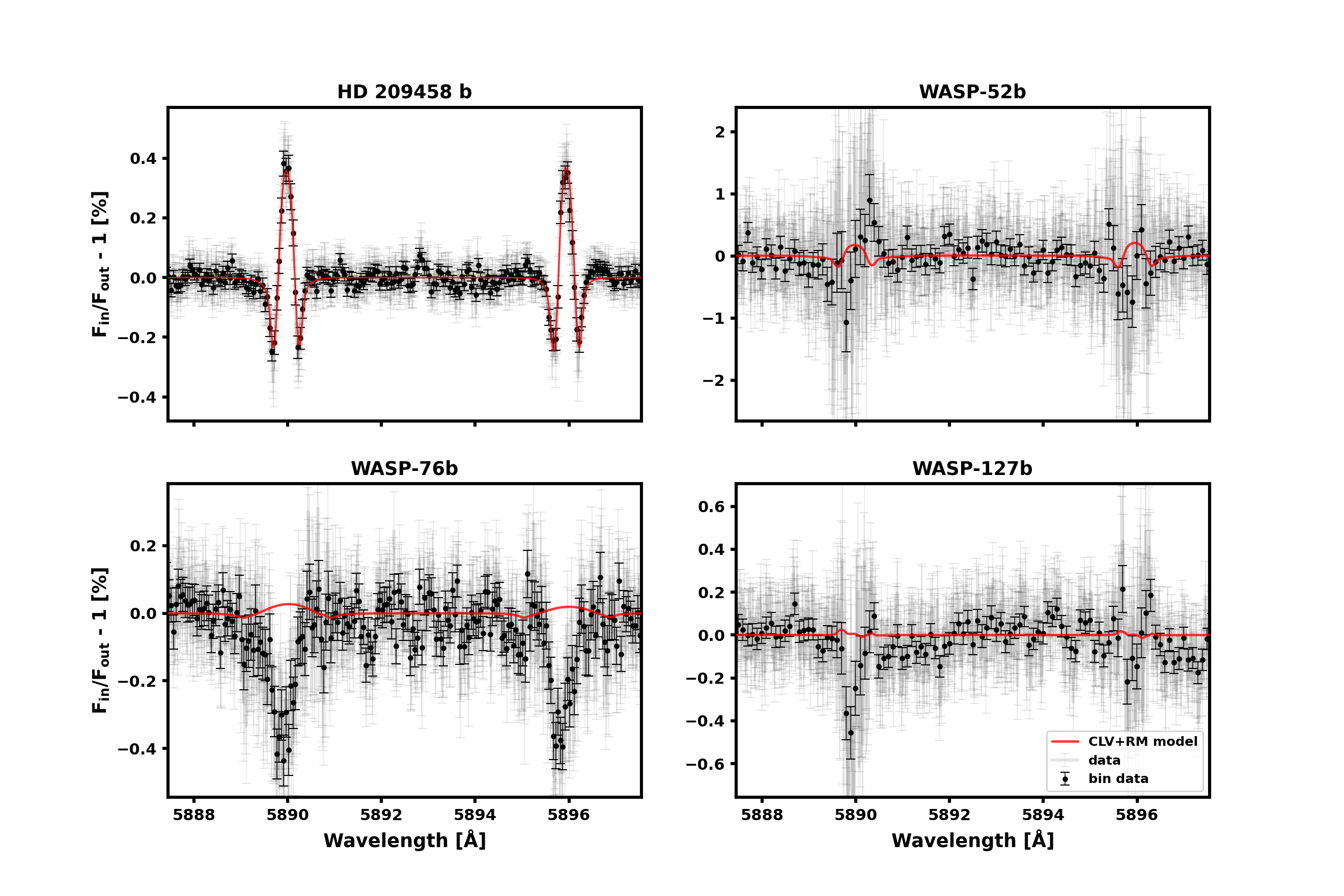}}
            
      \caption{Wiggles-corrected transmission spectra (in black) of the \ion{Na}{I} D lines with overplotted the CLV+RM model from 3D NLTE stellar spectra (in red) for: HD 209458 (upper left panel), WASP-52 (upper right panel), WASP-76 (middle left panel), WASP-127 (middle right panel).} 
              
         \label{fig: wigglescorrNa}
\end{figure*}
\begin{figure*}
\resizebox{\hsize}{!}
          {\includegraphics[width=5cm]{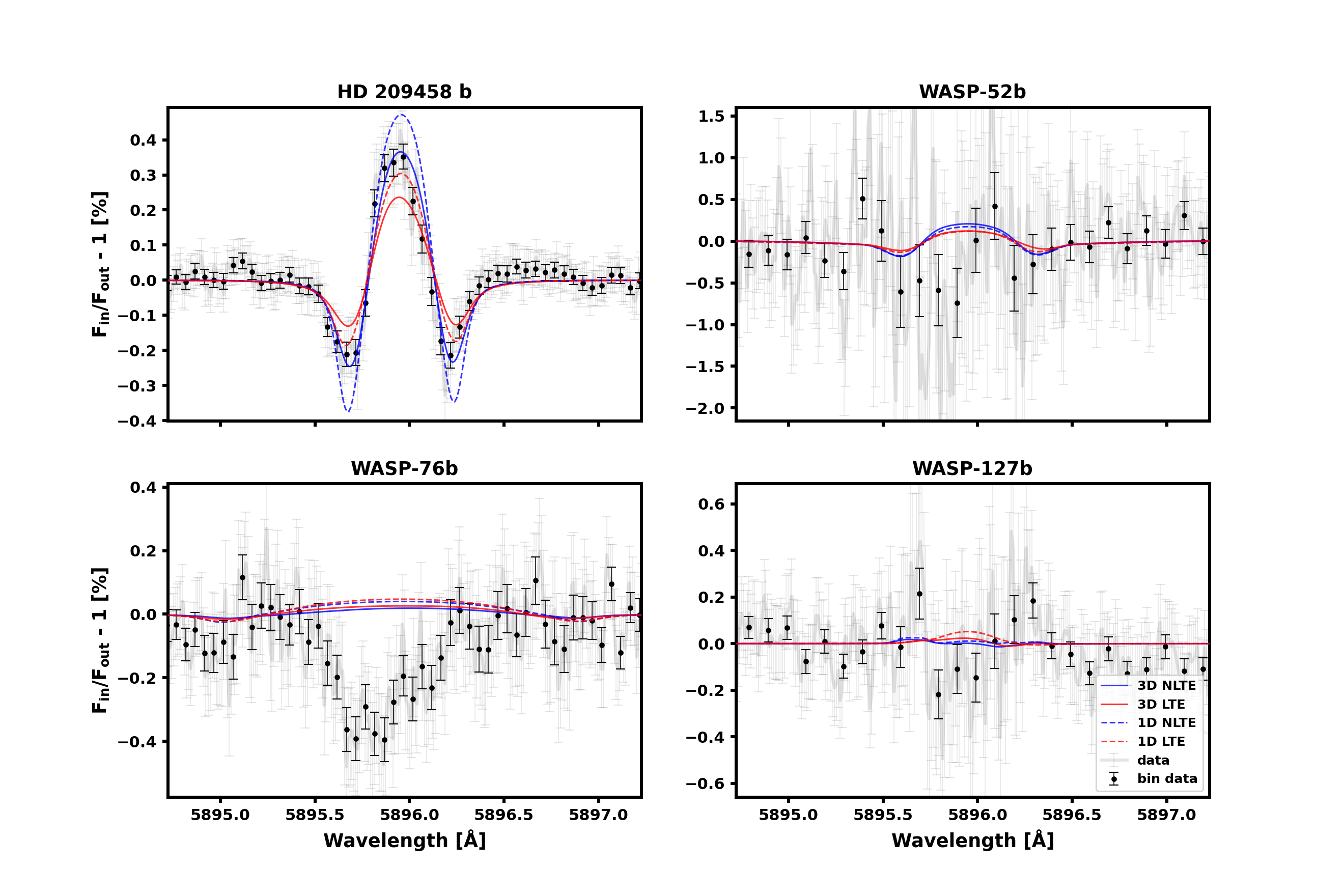}}
          \caption{Same as Fig. \ref{fig: wigglescorrNa} but zoomed in on the \ion{Na}{I} D$_1$ line with overplotted the CLV+RM model from different synthetic stellar spectra: 3D NLTE (solid blue line), 3D LTE (solid red line), 1D NLTE (dashed blue line), and 1D LTE (dashed red line).}
          \label{fig: wigglescorrNaALLD1}
\end{figure*}

\begin{figure*}
\resizebox{\hsize}{!}
          {\includegraphics[width=5cm]{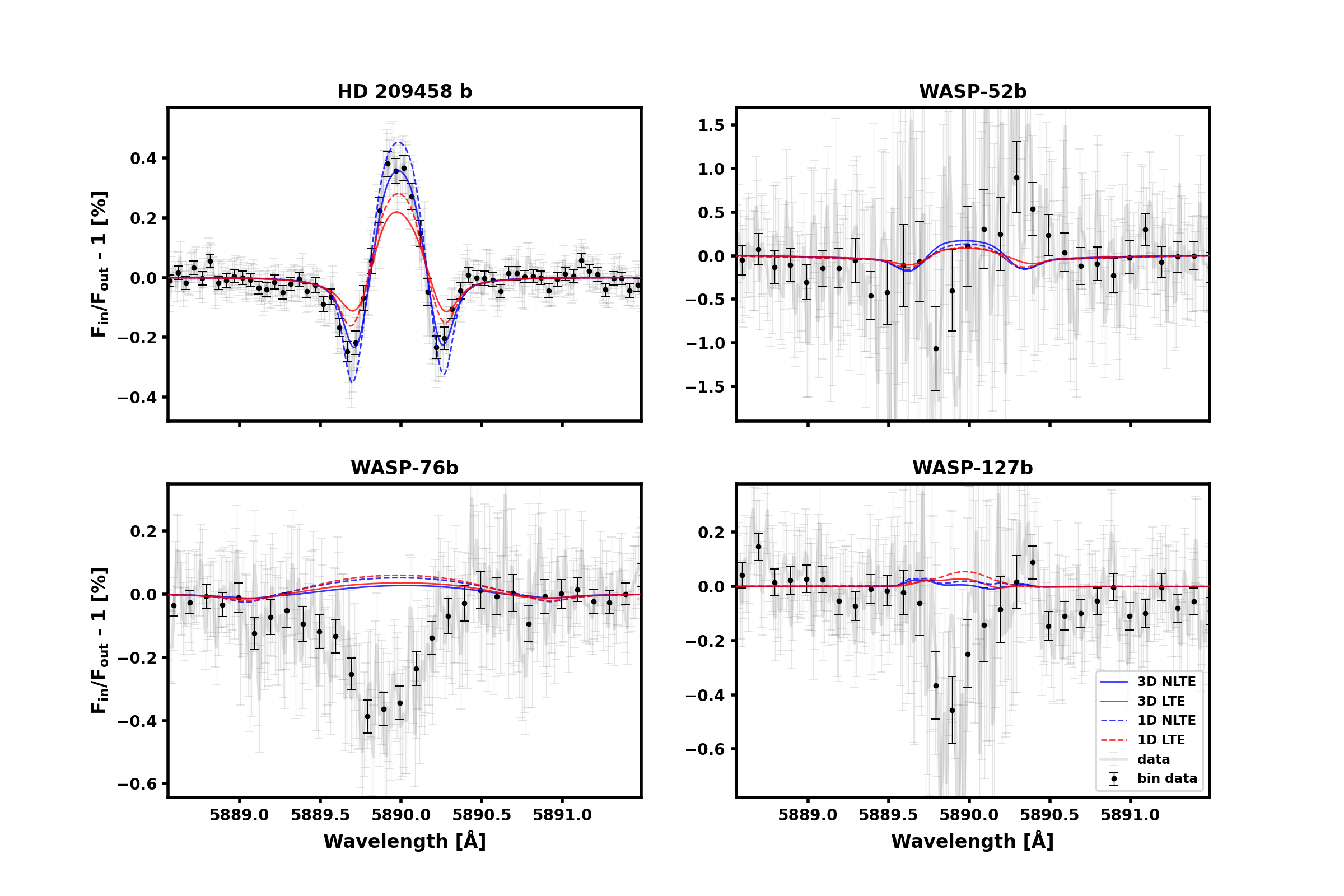}}
          \caption{Same as Fig. \ref{fig: wigglescorrNa} but zoomed in on the \ion{Na}{I} D$_2$ line with overplotted the CLV+RM model from different synthetic stellar spectra: 3D NLTE (solid blue line), 3D LTE (solid red line), 1D NLTE (dashed blue line), and 1D LTE (dashed red line).}
          \label{fig: wigglescorrNaALLD2}
\end{figure*}
\begin{figure}
    \includegraphics[width=0.5\textwidth]{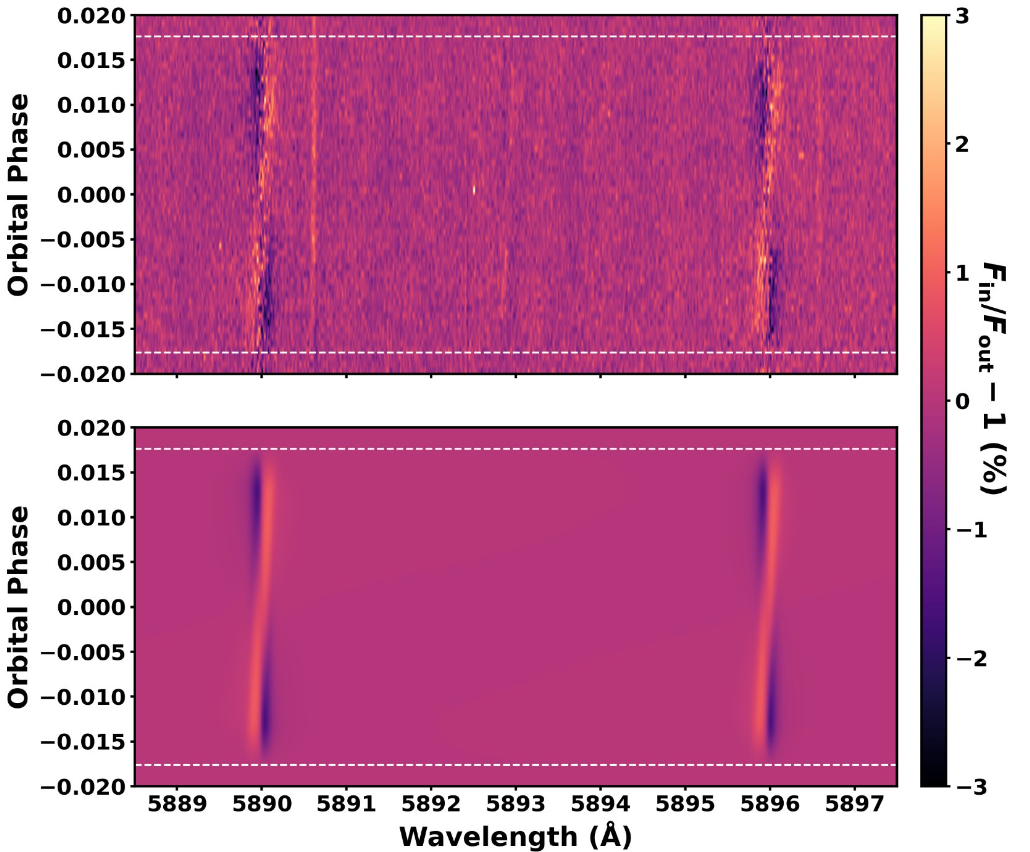} 
    \caption{Two-dimensional Map of HD 209458 during a transit by its planet (first night, T1), in the stellar rest frame. \textit{Top}: in- to out-of-transit flux observed by ESPRESSO on 2019-07-20. \textit{Bottom}: in- to out-of-transit flux produced with the best-fit 3D NLTE synthetic spectra. Dashed white lines represent the beginning and end of the transit.} 
    \label{fig: 2DMapHD209}
\end{figure}
\begin{figure*}
   \centering
   \includegraphics[width=\textwidth]{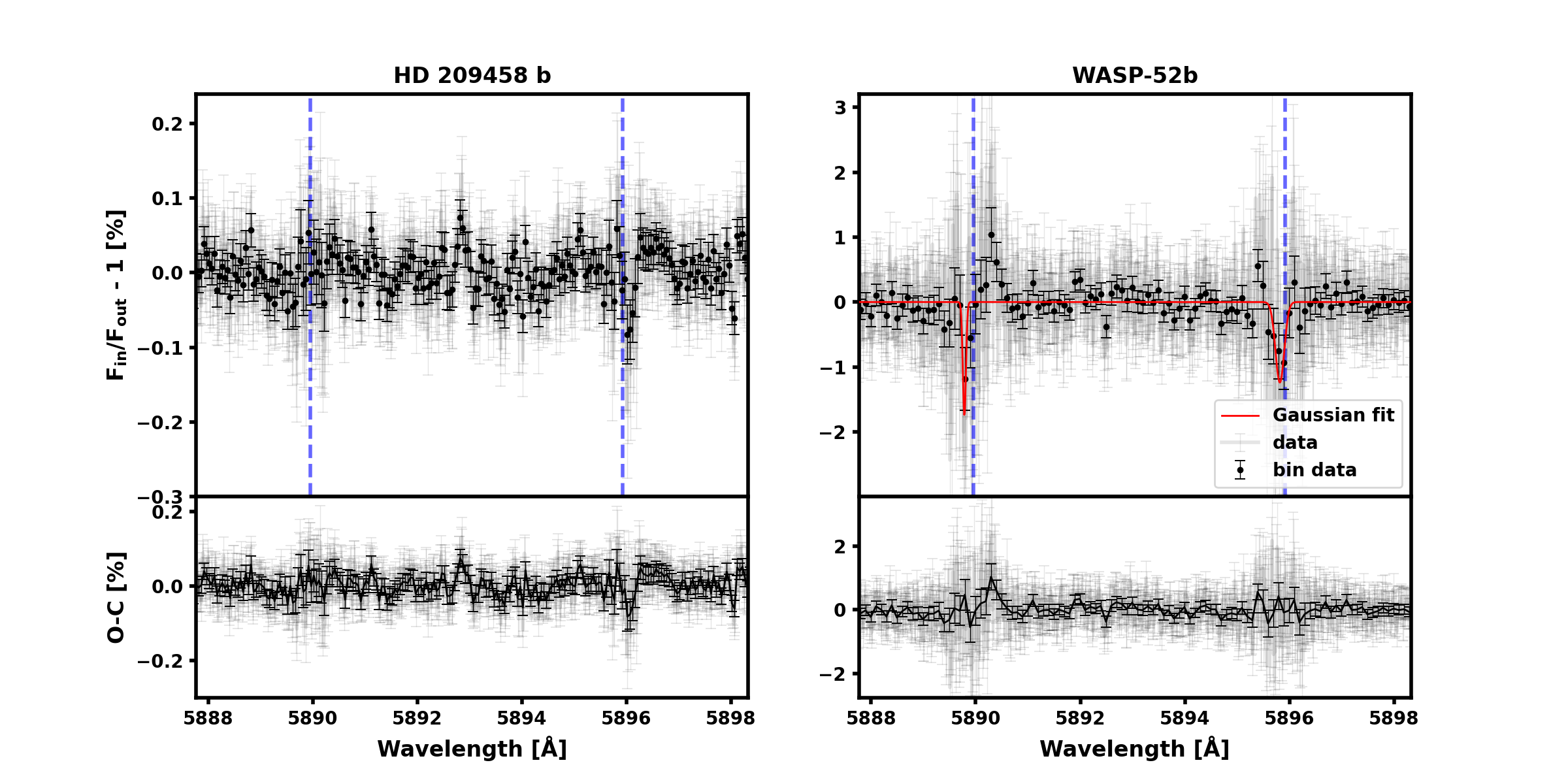}
   \includegraphics[width=\textwidth]{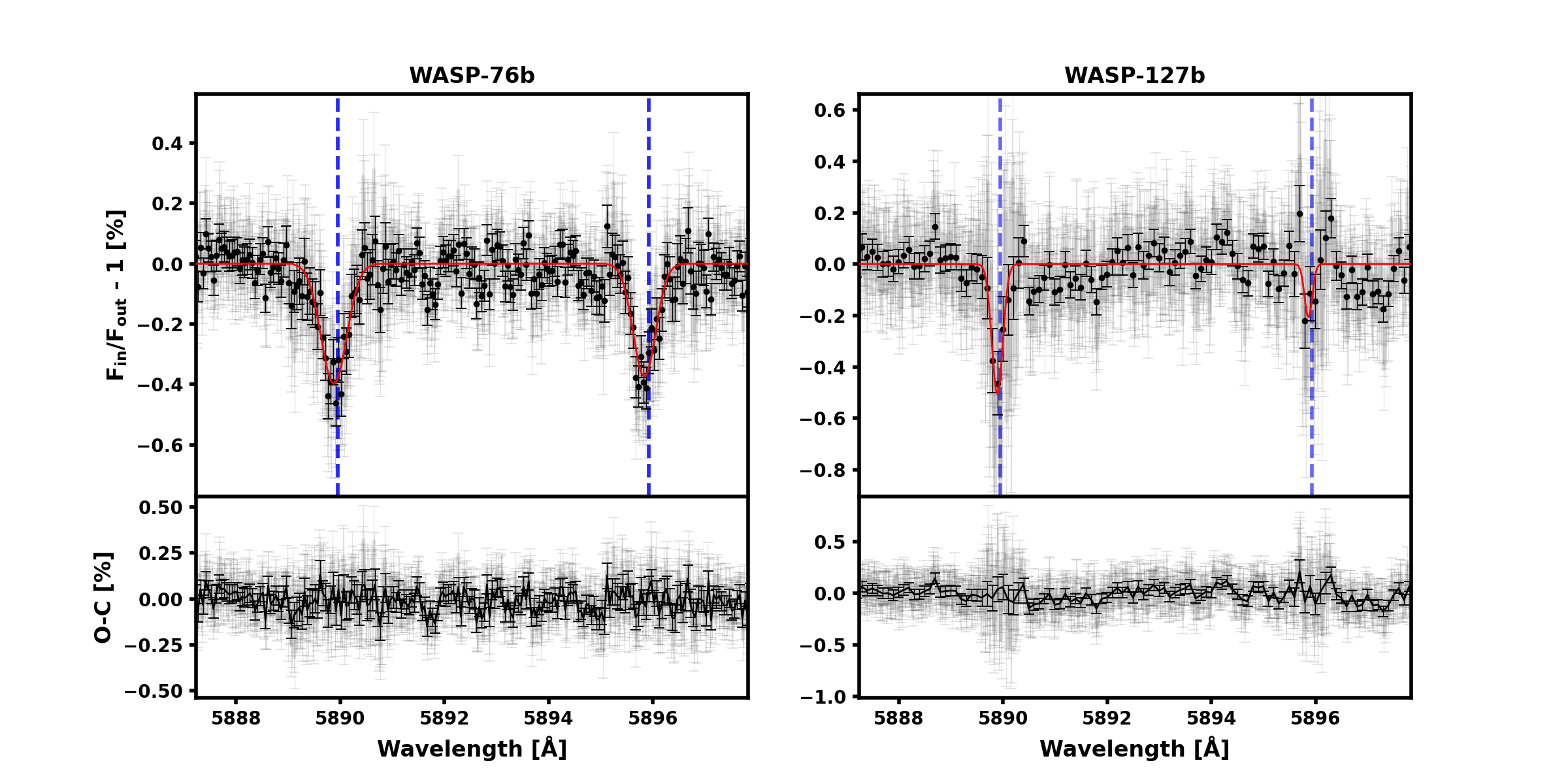}
   \caption{Transmission spectra corrected for the CLV+RM effect using 3D NLTE models, around the \ion{Na}{I} D lines. The gray dots represent the original data, whereas the black dots are the binned data (with a binning of 0.05\,\r{A} for HD 209458b and 0.1\,\r{A} for the others). \textit{Top}: The best Gaussian fit to the data is overplotted (red solid line). The rest-frame wavelength of each individual line is represented by a dashed blue vertical line. It is worth noticing the different scales of the y-axis. \textit{Bottom}: Residuals between the data and the Gaussian fit.}
   \label{fig: gaussfitNa}
\end{figure*}


The data reduction, including the sky correction, has been performed with the ESPRESSO data reduction pipeline, 
as specified in Table \ref{tab: ESPRESSOdata}. After that, for all the targets, we carried out the following steps to search for absorption signatures of \ion{Na}{I} in the doublet at 5890 and 5896\,\r{A}, originating from the upper layers of the planetary atmosphere. We extract the transmission spectrum as presented in previous studies (e.g., \citealt{Wyttenbach2015}; \citealt{Casasayas2017, Casasayas2021}; \citealt{YanHenning2018}; \citealt{Borsa2021}). 

\subsection{Telluric correction}\label{sec: telluric} 
The telluric emission lines like those from \ion{Na}{I} were removed in the sky subtraction step. On the other hand, the correction for telluric absorption lines coming from the Earth's atmosphere has been performed using the ESO software \texttt{Molecfit} (\citealt{Smette2015}; \citealt{Kaush2015}) through \texttt{ExoReflex} version 2.11.5. The code produces synthetic spectra of telluric H$_2$O and O$_2$ using a line-by-line radiative transfer model of Earth's atmospheric transmission. The correction is performed in the Earth rest frame and was successfully used in several works focusing on atmospheric composition using the ESPRESSO spectrograph (e.g., \citealt{Chen2020}; \citealt{Casasayas2021}; \citealt{Borsa2021}).  

\subsection{Determination of orbital phase}\label{sec: inout} 
For each night of observations, we calculate the orbital phase of each exposure based on literature ephemeris (such as transit duration, orbital period, and transit epoch), as reported in Table \ref{tab: systemparams1}. 
This allows us to determine the in- and out-of-transit exposures. The latter are defined as the frames that have zero overlap with the transit event. The fully in-transit exposures instead, are located for more than 50\% between the second and third contacts of the transit event. The other exposures are considered as either ingress or egress events. The exposures with a low S/N have been discarded as pointed out in Sect. \ref{sec: obs}. 

\subsection{RV fitting and creation of the Master Out}\label{sec: masterout}

A linear fit is performed on the radial velocities\footnote{The RV measurements are given as output of the ESPRESSO data reduction pipeline.} (RV) of the out-of-transit frames, in order to determine the semi-amplitude of the stellar radial velocity signal ($K_\star$, see Table \ref{tab: systemparams1}).
The in-transit exposures are discarded for this determination since they contain the RM component. 

At this point, all the spectra are shifted to the stellar rest frame (SRF) using the stellar RV-fitted model. The systemic velocity ($v_\mathrm{sys}$) is corrected for each night as well.

All the spectra are then normalized using a third-order polynomial function 
to fit the continuum variations between every individual spectrum. Indeed, the variations in the Earth's atmosphere (e.g., airmass) during the night, make the flux vary between each exposure.
Then, the normalized out-of-transit spectra are averaged together to create the Master Out spectrum, using the inverse square of flux uncertainties as the weight. 
All the non-discarded out-of-transit observations, as specified in Sect. \ref{sec: HD209intro} to \ref{sec: WASP127intro}, are used to generate the Master Out, resulting in an average relative uncertainty in the final flux ranging from 0.1\% for HD 209458 to 0.8\% for WASP-52.
This procedure has been repeated for each night and the resulting spectra for all the targets are shown in Fig. \ref{fig: FoutNa}. In this figure, the best-fit synthetic spectrum produced with 3D NLTE models, as described in Sect. \ref{sec: synthspectra}, is overplotted.

\subsection{Correction for the center-to-limb variation and Rossiter-McLaughlin effect}\label{sec: CLVRM}
We model the combined effect of the CLV and RM on the \ion{Na}{I} D line profiles using the best-fit synthetic stellar spectra as described in Sect. \ref{sec: synthspectra}. We consider the 3D NLTE, 3D LTE, 1D NLTE, and 1D LTE stellar synthetic spectra.

Following the approach of \citet{Casasayas2019} and \citet{Morello2022}, the sky-projected stellar disk is represented as a grid of $300 \times 300$ squared cells. Each cell has a specific $\mu$-angle, radial velocity ($v_\mathrm{\mu}$), and intensity spectrum ($I_\mu$) depending on its position on the disk. 
The intensity spectra are interpolated from the best-fit synthetic spectra at 21 $\mu$-angles, ranging from the disk center ($\mu=1.0$) to near the edge ($\mu=0.001$), where $\mu=0.001$ was chosen instead of 0.0 to avoid numerical errors.

The spectra for cells at intermediate $\mu$-angles is obtained via linear interpolation. 
Each cell is associated with Cartesian coordinates ($x_i, y_i$), assuming a circular stellar disk with a radius $R_\star =1$, centered on the origin. Therefore, for the $i$-th cell the $\mu$-angle is calculated as:
\begin{equation}
    \mu_i= \sqrt{1 - x^2_i - y^2_i} \ ,
\end{equation}
The local radial velocity ($v_i$) for each cell is determined by the sky-projected obliquity ($\lambda$) and the stellar rotational velocity ($v \sin i_\star$), using the following equation:
\begin{equation}
    v_i = v \sin i_\star (y_i \sin \lambda -x_i \cos \lambda) \ .
\end{equation}
The values of $\lambda$ and $v \sin i_\star$ for the four targets of this work are taken from literature and are reported in Table \ref{tab: systemparams1}.
These values can be retrieved by modeling the RM effect on the in-transit spectra, for example through the reloaded RM technique (\citealt{Cegla2016}) adopted in CB21 to analyze the HD 209458b data. 

In these simulations, we assume that the star rotates as a solid body, meaning that differential rotation is not considered. At each orbital phase, the stellar spectrum is then obtained by summing the Doppler-shifted spectra from all stellar disk cells not obscured by the transiting planet. 
The resulting CLV+RM models in the transmission spectra using 3D NLTE synthetic spectra around the \ion{Na}{I} D lines of each target are shown in Fig. \ref{fig: wigglescorrNa}. The comparisons between CLV+RM models obtained using different synthetic spectra, that is 1D LTE, 1D NLTE, 3D LTE and 3D NLTE, are shown in Fig. \ref{fig: wigglescorrNaALLD1} and \ref{fig: wigglescorrNaALLD2}, for the \ion{Na}{I} D$_1$ and D$_2$, respectively. Finally, in Fig. \ref{fig: transLCs075} and \ref{fig: transLCs04} the CLV+RM models for the transmission LCs are displayed in Appendix.

\subsection{Creation of the transmission spectrum}\label{sec: transspectr}
To obtain the transmission spectrum of each exposure in the stellar rest frame, each in-transit spectrum in the SRF ($F_{i,\mathrm{in}}$) is divided by the Master Out. As the planet moves across the stellar disk during a transit, it gradually blocks the light from regions with varying velocities and CLV. In the case of an aligned orbit, the planet initially covers the blueshifted side of the star at ingress, followed by the redshifted side near egress, causing distortions in the spectral line profiles due to the star’s rotation. This phenomenon is referred to as the Doppler shadow (\citealt{CollierCameron2010}). The latter is clearly visible in the two-dimensional map in Fig. \ref{fig: 2DMapHD209}, where the ratio of the in- to out-of-transit flux observed for each phase of HD 209458b is displayed, together with the corresponding CLV+RM model from the 3D NLTE synthetic spectra.

All the residual spectra are then shifted to the planet rest frame (PRF) by computing the planet velocity at each phase $\phi$ through the following equation:
\begin{equation}
    v_\mathrm{p}= K_\mathrm{p} \ \sin (2\pi\phi) \ , 
    \label{eq: vp}
\end{equation}
where $K_\mathrm{p}$ is the planet RV semi-amplitude obtained from transit parameters such as the semi-major axis $a$, the orbital inclination $i_\mathrm{p}$, and the orbital period $P_\mathrm{orb}$ as follows, assuming zero eccentricity:
\begin{equation}
    K_\mathrm{p}= \frac{2 \pi a}{P_\mathrm{orb}} \sin (i_\mathrm{p}) \ .
    \label{eq: Kp}
\end{equation}
The $K_\mathrm{p}$ calculated with this equation for each target is reported in Table \ref{tab: systemparams1}. 
Finally, to compute the transmission spectrum, all the in-transit residuals are combined with an arithmetic average.

For better visualization of the resulting transmission spectra, they are binned in wavelength to either 0.05\,\r{A} (for HD 209458b) or 0.10\,\r{A} (for WASP-52b, WASP-76b, and WASP-127b), as can be seen in Fig. \ref{fig: wigglescorrNa}.

An issue with ESPRESSO data that has been highlighted in several previous studies (e.g., \citealt{Casasayas2021}; \citealt{Tabernero2021}), is the interference sinusoidal pattern, "wiggles", arising when dividing spectra taken at different times during the same night, that is in transmission spectra. These wiggles are due to non-flat-fielded interference occurring in some optical elements of the Coud{e'} train. In this work, the ESPRESSO wiggles are corrected by fitting a third-order spline to the combined transmission spectrum of each night, as previously performed by \citet{Zewen2023}. An example of this correction is shown in Fig. \ref{fig: wiggleexample} in Appendix. 

After the wiggle-pattern correction, the CLV+RM model is subtracted from the data as well, in order to obtain the final transmission spectrum. Transmission spectra from different transits are co-added using a weighted average. 
Finally, a Gaussian fit is performed on the absorption features (if any) corresponding to the \ion{Na}{I} D lines. The corrected transmission spectra are shown in Fig. \ref{fig: gaussfitNa}. 
The absorption depth inferred from the Gaussian fit is reported in Table \ref{tab: depths}, in comparison with measurements from previous works that used the same datasets. Other properties of the line fit (i.e., FWHM and Doppler shift) are reported in Table \ref{tab: winds} in Appendix.

\subsection{Creation of the transmission light curve}\label{sec: translc}
As done in previous studies on the \ion{Na}{I} D lines (e.g., \citealt{Yan2017}; \citealt{Casasayas2020, Casasayas2021}), considering the transmission spectra in the PRF, the transmission LCs are obtained by fixing a passband of a given width $\Delta \lambda =0.75$\,\r{A} or 0.4\,\r{A} around the center of the line to analyze, and then by calculating the flux inside the passband using trapezoidal integration. 
In order to perform this calculation, we use tools from \texttt{ExoTETHyS}\footnote{\url{https://github.com/ucl-exoplanets/exotethys}} (\citealt{Morello2020a, Morello2020b,Morello2021}), following the same procedure detailed in \citet{Morello2022}. As in CB21, the LCs of different nights are combined by sorting the values measured at different time stamps in chronological order with respect to the center of the transit. In this way, an interpolation to a common time axis can be avoided. For visualization purposes, the combined data are then binned to 0.001 in orbital phase.
The transmission LCs of the average of the two \ion{Na}{I} D lines of HD 209458b are displayed in Fig. \ref{fig: HD209results}, while for the other targets are shown in Appendix in Fig. \ref{fig: transLCs075} and \ref{fig: transLCs04}, for the 0.75\,\r{A} and 0.4\,\r{A} passband, respectively.

\section{Results}\label{sec: analysis}
\begin{figure}
    \centering
    \includegraphics[width=0.5\textwidth]{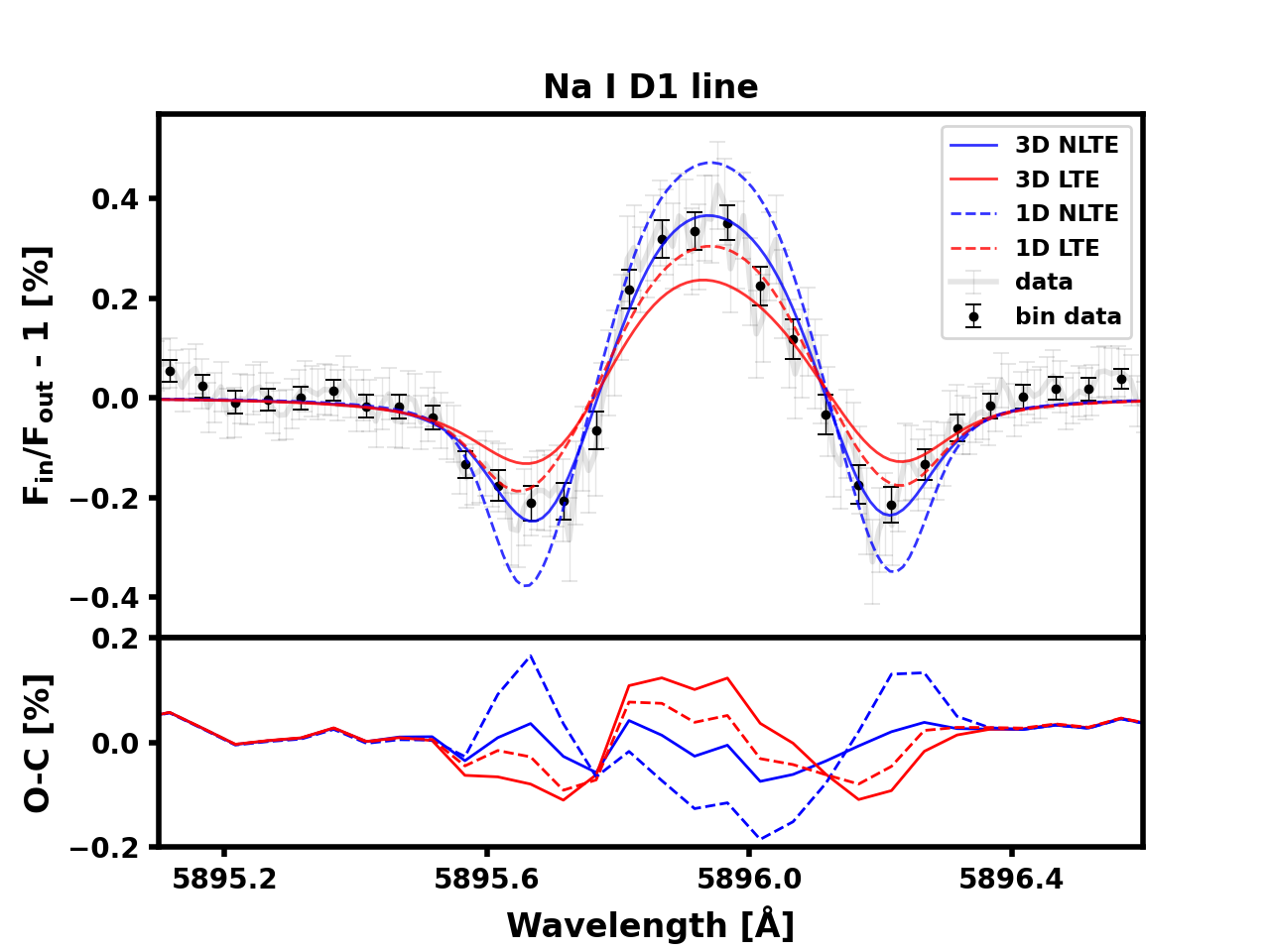} 
    \caption{Transmission spectrum of HD 209458b centered on the \ion{Na}{I} D$_1$ line in comparison with models of the CLV+RM effect from synthetic stellar spectra, colored as in the figure's legend. In the bottom panel, the difference between the binned data and the models is shown.} 
    \label{fig: HD209NaD1}
\end{figure}
\begin{table*}
\centering 
\small 
\caption{Measured depths of the Gaussian fits to the \ion{Na}{I} D lines in the transmission spectra for all targets in this work, after correcting for the CLV+RM effect with 3D NLTE models.}\label{tab: depths} 
\begin{tabular}{l c c c c c c}
\\ \hline \hline 
\noalign{\smallskip}
Planet  & Transit & \multicolumn{2}{c}{This work} & \multicolumn{2}{c}{Previous work}  & Reference \\ 
 & & D$_1$ (\%) & D$_2$ (\%) & D$_1$ (\%) & D$_2$ (\%) & \\ \hline
 \noalign{\smallskip}
HD 209458b & & \multicolumn{2}{c}{$0.000 \pm 0.026^{(a)}$} & - & - & \citet{Casasayas2021}\\
WASP-52b & T1 & $1.38 \pm 0.35$ & $2.20 \pm 0.31$ & - & - & - \\
 & T2 & $1.31 \pm 0.46$ & $0.92 \pm 0.44$ & - & - & -\\
 & T3 & $1.26 \pm 0.34$ & $2.94 \pm 0.23$ & - & -  & -\\
 & coadded & $1.25 \pm 0.19$ & $1.75 \pm 0.32$ & $1.09 \pm 0.16$ & $1.31 \pm 0.13$ & \citet{Chen2020} \\
WASP-76b & T1 & $0.400 \pm 0.032$ & $0.443 \pm 0.020$ & $0.385 \pm 0.051$ & $0.449 \pm 0.049$ & \citet{Tabernero2021}\\
 & T2 & $0.326 \pm 0.029$ & $0.316 \pm 0.013$ & $0.294 \pm 0.042$ & $0.246 \pm 0.037$ & " \\
 & coadded & $0.376 \pm 0.021$ & $0.389 \pm 0.022$ & $0.331 \pm 0.032$ & $0.319 \pm 0.029$ & " \\ 
WASP-127b & T1 & $0.46 \pm 0.08$ & $0.64 \pm 0.03$ & - & - & - \\ 
& T2 & $0.18 \pm 0.06$ & - & - & - & - \\
& coadded & $0.21 \pm 0.05$ & $0.51 \pm 0.04$ & \multicolumn{2}{c}{$0.34 \pm 0.04$ (mean of D$_1$ and D$_2$)} &  \citet{Allart2020} \\ \hline  
\end{tabular}\\
\begin{flushleft}
\textbf{Notes:}\\
The "–" symbol is used to denote no previous measurements of individual lines via Gaussian fits.\\
$^{(a)}$ This value is an estimate of the noise level in the \ion{Na}{I} D region of the transmission spectrum. Therefore, the upper limit for a $3\sigma$ detection would be about 0.08\%.
\end{flushleft}
\end{table*}
We apply the different synthetic models described in Sect. \ref{sec: synthspectra} to correct for the stellar spectrum effect (CLV+RM model) imprinted on ESPRESSO transmission data of four giant exoplanets orbiting around relatively bright host stars. We compare the four stellar models computed with different assumptions, that is 1D LTE, 1D NLTE, 3D LTE, and 3D NLTE, to the transmission spectrum of the targets in Sect. \ref{sec: resultsspectra}. The transmission light curve of HD 209458b is also analyzed in Sect. \ref{sec: resultsLC}.

Moreover, we discuss the CLV effects over the grid of 3D NLTE synthetic stellar spectra in Sect. \ref{sec: resultsgrid}. We also simulate how the CLV+RM effect of the HD 209458 system varies within the parameter space of the stellar grid in Sect. \ref{sec: HD209stars}.

\subsection{Transmission spectra}\label{sec: resultsspectra}
As already shown in several works (e.g., \citealt{Czesla2015}; \citealt{Casasayas2020}), depending on the geometry of the star-planet system, the amplitude of the CLV+RM effect can be as strong as the absorption feature itself. One of the main parameters affecting the shape of the CLV+RM effect is $\lambda$, the sky-projected angle. 
For aligned orbits, such as those of HD 209458b and WASP-52b (see Table \ref{tab: systemparams1}), the amplitude of the stellar CLV on the planetary transmission spectrum and LCs can be as large as 0.5\% (e.g., \citealt{Yan2017}). This is indeed the case for the HD 209458 and the WASP-52 systems. Specifically, for HD 209458b the difference between the adopted stellar model is significant, and indeed a large scatter up to $\pm$ 0.2\% for the 1D models is observed when computing the residuals between the data and the models: $+0.2\%$ for the 1D NLTE, and about $-0.2\%$ for the 1D LTE, respectively. This is illustrated in Fig. \ref{fig: HD209NaD1}, where the transmission spectrum is zoomed-in on the \ion{Na}{I} D$_1$ line to better visualize the differences between the synthetic spectra. Different stellar models result in varying amplitudes of the line deformations. 
In Fig. \ref{fig: HD209NaD1}, it can be clearly seen that only the 3D NLTE model results in an extremely good fit of the observed transmission spectrum of the \ion{Na}{I} D$_1$ line. On the other hand, the 1D LTE and NLTE models result in an underestimate and overestimate, respectively, of the observed amplitude in the line core. The same behavior is observed for the \ion{Na}{I} D$_2$ line, as shown in the top left panel of Fig. \ref{fig: wigglescorrNaALLD2}.
Specifically, the residuals obtained with the 1D NLTE model show a dip in the line core which can be mistaken for a Na absorption feature corresponding to an absorption depth of about 0.2\%, if this stellar model is adopted. 
Then, if a 3D stellar atmosphere is used in the LTE assumption, the resulting CLV+RM model produces a curve with the smallest amplitude (of 0.23\% at maximum) compared to all the other stellar models, which translates into a peak in the line core of about +0.2\% in the corresponding residuals. 
With the 3D NLTE models, instead, the CLV+RM curve shows a peak at about 0.4\% at the line center and a minimum of about $-0.25$\% in the line wings, perfectly matching the data. Therefore, we obtain essentially flat residuals when subtracting the 3D NLTE model from the data, indicating that no additional Na absorption is required to explain the behavior of the \ion{Na}{I} D lines in the transmission spectrum of HD 209458b. This supports the results of CB21, namely that no Na is detected in the atmosphere of this planet at the layers probed by high-resolution spectroscopy.
However, this result does not necessarily contradict the many Na detections at low-resolution. Indeed Na could still be present in small quantities, colder than equilibrium temperature and/or deeper atmospheric layers without significantly affecting the transmission spectrum at high-resolution, as shown by \citet{Morello2022}.

For the other planets, namely the WASP systems, the average S/N is lower and the error bars on the data points are slightly larger compared to HD 209458b, making it more challenging to differentiate between various stellar models. Despite these challenges, there remain differences between models, although the detection and measurement of Na absorption can be achieved even without correcting for the CLV+RM effect, albeit with a reduced confidence level. After correcting for the CLV+RM effect using a 3D NLTE model, we measure an absorption depth consistent with previous literature studies that either did not apply this correction or used a 1D model. The results are summarized in Table \ref{tab: depths}.

For WASP-52b, the amplitude of the 3D NLTE curve is the largest among the models, reaching a peak of about 0.2\% at the line center. The differences with the other stellar models are less than 0.1\%, as shown in Fig. \ref{fig: wigglescorrNaALLD1} and \ref{fig: wigglescorrNaALLD2}. Given that the error bars on the binned data range between 0.13\% and 0.47\%, it is not possible to distinguish the results among the models (see Table \ref{tab: Nadetmodels} in Appendix).  
The planetary absorption obtained after correcting for the CLV+RM effect with the 3D NLTE model is shown in the top right panel of Fig. \ref{fig: gaussfitNa}, where the Na features have been fitted with Gaussians. The absorption depths measured after correcting for the CLV+RM effect with 3D NLTE models are larger than those measured by \citet{Chen2020}, who used 1D LTE SME models. 
Specifically, the absorption depth of the \ion{Na}{I} D$_1$ is larger by about 14\%, whereas the D$_2$ line depth increases by about 33\%, as reported in Table \ref{tab: depths}. Without the CLV+RM correction, we measure smaller absorption depths, even closer to the results of \citet{Chen2020}. 
Our results are generally consistent with theirs, except for the Doppler shift in wavelength. While they do not observe this shift, our Gaussian fits indicate a blueshift corresponding to a wind velocity of approximately $-8.0$\,km\,s$^{-1}$. 
Additionally, no broadening in the absorption line wings is observed.

With the continuous improvements of high-resolution spectrographs, higher S/N will likely be achieved in the near future, resulting in lower error bars similar to those for HD 209458b. Consequently, adopting the most accurate stellar model will be crucial for the correct characterization of exoplanetary atmospheres.

For WASP-76b, the error bars on the binned data are of the same order of magnitude as HD 209458b, ranging between 0.04\% and 0.08\%. However, due to the significantly misaligned orbit ($\lambda=61.28$\,deg), the amplitude of the CLV+RM effect becomes quite small. Specifically, it reaches a maximum of 0.03\% for the 3D NLTE model and 0.06\% for the 1D LTE model, which exhibits the largest amplitude in this instance. Moreover, in this case, the Na absorption is so evident in the transmission spectrum that it can be detected even without correcting for the CLV+RM effect (see Fig. \ref{fig: wigglescorrNa}), which should not make a significant difference, given its amplitude. However, after correcting the transmission spectra for the CLV+RM effect with the 3D NLTE model and fitting the \ion{Na}{I} D lines, we find a Na absorption depth larger than the values reported by \citet{Tabernero2021} by about 14\% and 22\% for the D$_1$ and D$_2$ lines, respectively (see Table \ref{tab: depths}). Furthermore, we are able to retrieve the Na absorption with a larger confidence level, that is 18$\sigma$ instead of the 9.2$\sigma$ of \citet{Tabernero2021}. Therefore, we conclude that, given the WASP-76b absorption depth is about 0.4\% (much smaller than in the WASP-52b case), the correction for the CLV+RM effect does indeed make a difference by improving the detection confidence level, even though the amplitude of the effect is extremely small. 
Moreover, the absorption features we observe are slightly broadened and blueshifted of about $-4.3 \pm 0.5$\,km\,s$^{-1}$, indicating the presence of winds in the planet's atmosphere, which is consistent with the results of \citet{Tabernero2021}.

For WASP-127b, a system with a heavily misaligned and retrograde orbit (i.e., $\lambda=-128.41$\,deg), all the CLV+RM models result in an essentially flat curve with a maximum amplitude of about 0.03\%, except for the 1D LTE model that reaches a peak of 0.05\%. However, given the spread of the data points and the error bars being of the same order of magnitude as the CLV+RM effect, namely within 0.04--0.13\%, correcting for this effect does not appear to be important in this system. Indeed, after correcting for the CLV+RM effect with the 3D NLTE model, we retrieve an absorption depth averaged between the two \ion{Na}{I} D lines of $0.35 \pm 0.04$\,\%, which is perfectly in agreement with the value measured by \citet{Allart2020}, who did not perform any correction. We also observe an average Doppler shift of $-3.2 \pm 0.4$\,km\,s$^{-1}$, indicating the presence of winds in the planetary atmosphere, as previously reported by \citet{Allart2020}.
In line with their results, we also find a significant asymmetry between the absorption depths of the \ion{Na}{I} D lines, with a line ratio of D$_2/$D$_1=2.4$. 
Previous studies investigated possible explanations for a D$_2/$D$_1$ line ratio differing from unity, suggesting a variety of scenarios: ranging from stellar activity to variability in the chemical composition or velocity distribution of the \ion{Na}{I} atoms, as well as whether the absorption occurs in an optically thin or thick regime (\citealt{GebekOza2020}).

Our results suggest that for extremely misaligned orbit, like in the WASP-127 system, there is no need for correcting for the CLV+RM effect, since it does not make any significant impact on the detection and measurement of \ion{Na}{I} absorption, if present.

\begin{figure}
    \includegraphics[width=0.53\textwidth]{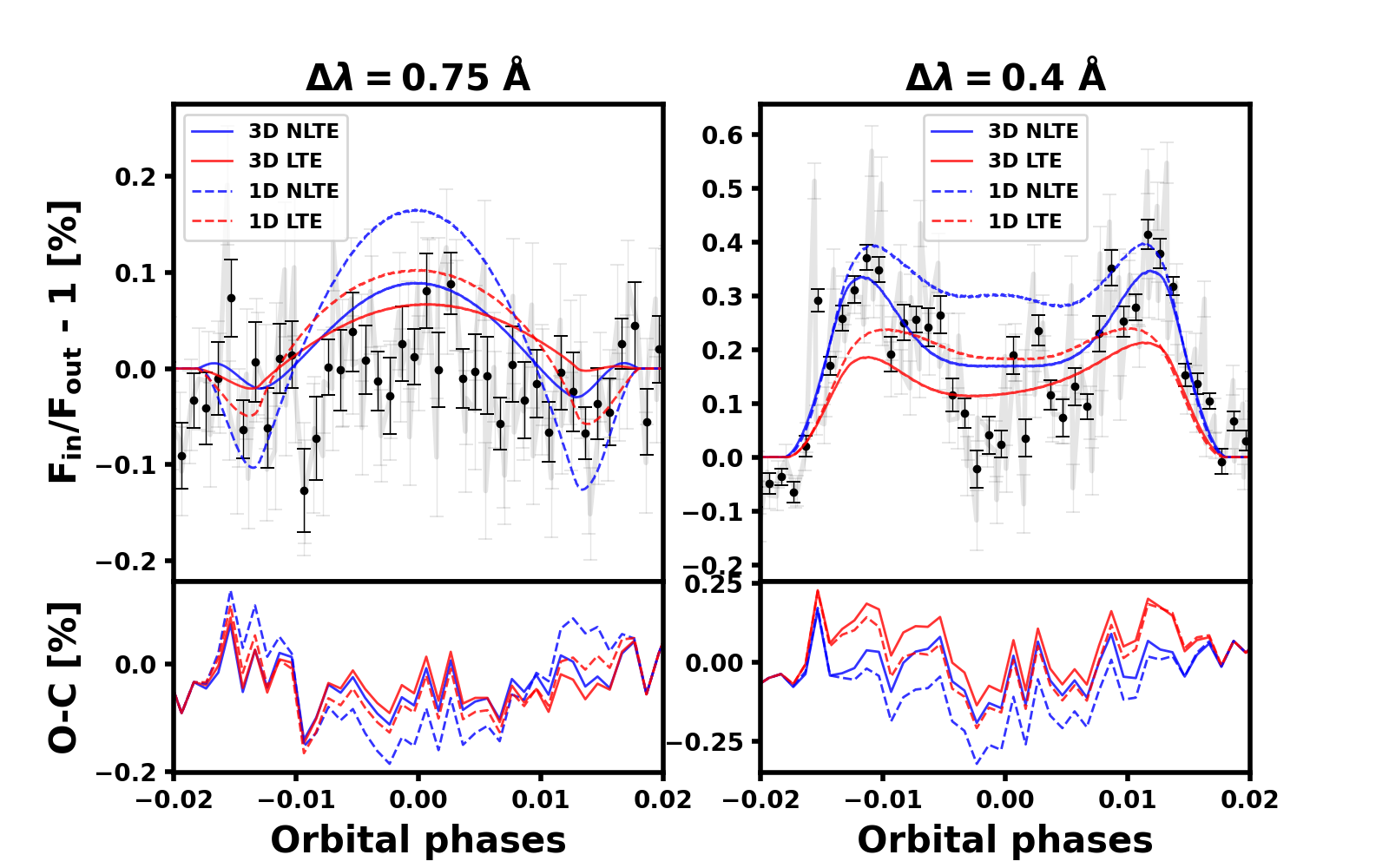} 
    \caption{Transmission LCs of HD 209458b in comparison with models of the CLV+RM effect from synthetic stellar spectra, colored as in the figure's legend. \textit{Left}: transmission LC computed in a passband of 0.75\,\r{A} around the \ion{Na}{I} D lines, averaged together, and binned to 0.001 in orbital phases for better visualization. \textit{Right}: transmission LC in a passband of 0.4\,\r{A}. In the bottom panels, the difference between the binned data and the models is shown.} 
    \label{fig: HD209results}
\end{figure}
\begin{figure*}
    \resizebox{\hsize}{!}
            {\includegraphics[width=15cm]{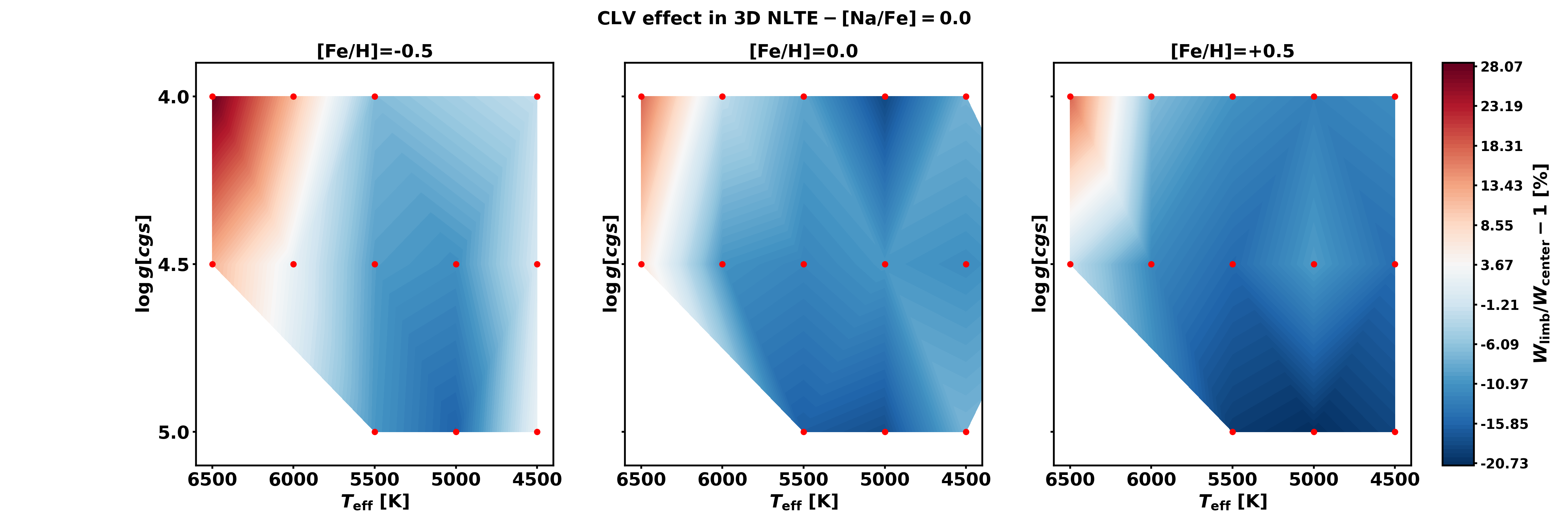}}
            
      \caption{The CLV effect across the \textsc{Stagger}-grid for FGK-type stars. The panels show the ratio of the combined equivalent width of the \ion{Na}{I} D lines from 3D NLTE synthetic spectra at the limb ($\mu=0.1$) and at the disk center ($\mu=1.0$). The red dots indicate the grid nodes where the synthetic spectra have been computed.} 
         \label{fig: CLV effect}
\end{figure*}
\begin{figure*}
\resizebox{\hsize}{!}
          {\includegraphics[width=5cm]{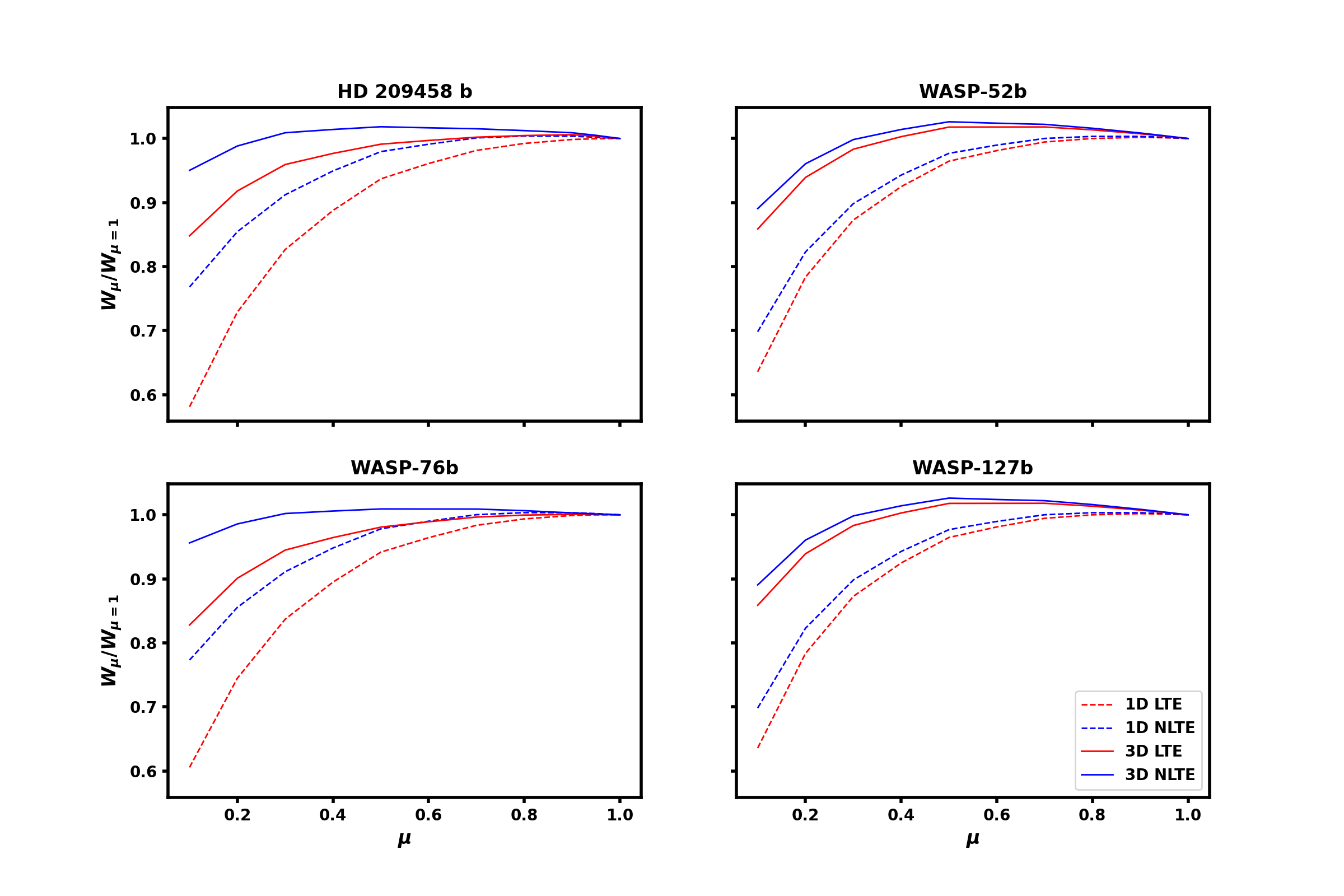}}
          \caption{Center-to-limb variation of the equivalent width ($W_\lambda$) of the combined \ion{Na}{I} D lines for the best-fit synthetic spectra of the four targets, colored as in the figure's legend.}
          \label{fig: CLVEWsbestfit}
\end{figure*}
\begin{figure*}
    \resizebox{\hsize}{!}
            {\includegraphics[width=15cm]{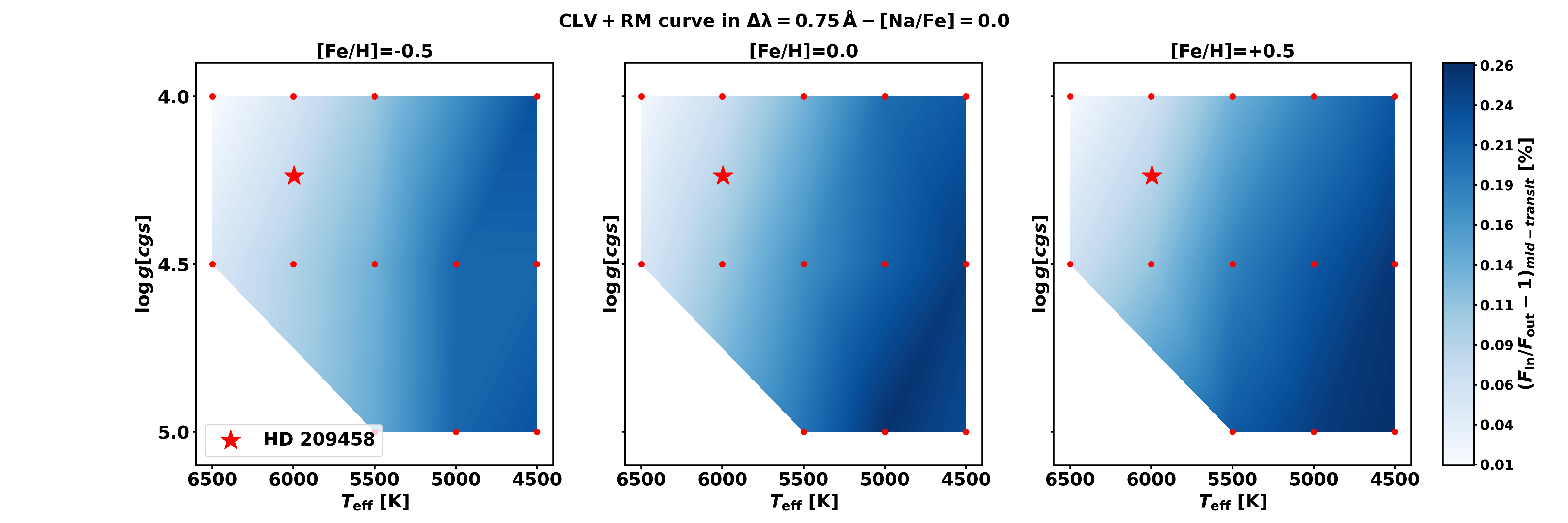}}
            
      \caption{Contour diagram of the mid-transit value of the CLV+RM model of the transmission light curve in a 0.75\,\r{A} band, for a planet with the same planetary and orbital properties of HD 209458b but orbiting FGK-type stars from the \textsc{Stagger}-grid. The red dots indicate the grid nodes where the synthetic spectra have been computed. The red star symbol indicates where the true HD\,209458 star lies in the grid.}          
         \label{fig: HD209_LC75_grid}
\end{figure*}
\subsection{Transmission light curves}\label{sec: resultsLC}
As previously observed in other studies, the shape of the transmission LCs varies depending on the selected bandwidth: narrower bandwidths result in stronger LC amplitudes (e.g. \citealt{Yan2017}; \citealt{Canocchi2024}).
Therefore, we selected the narrow bandwidths of 0.75\,\r{A} and 0.4\,\r{A} to maximize the LC amplitude to compare our models with.


The transmission LCs in bandwidths of 0.75\,\r{A} and 0.4\,\r{A} for HD 209458b are shown in Fig. \ref{fig: HD209results}, in the left and right panels, respectively. In both cases, the 1D NLTE model overestimates the amplitude at mid-transit.

In the 0.75\,\r{A} LC, it is difficult to distinguish the best model, as all of them are within the error bars of the data points, except for the 1D NLTE one that clearly overestimates the data in the mid-transit part.

On the other hand, in the 0.4\,\r{A} case, the LTE models, both 1D and 3D, underestimate the ingress and egress parts of the transit. A better match is obtained from the 3D NLTE model, but it is still not perfect in the mid-transit part, as it does not go deep enough. To completely rule out the presence of \ion{Na}{I}, we would expect the CLV+RM model to perfectly fit the observed data, as seen in the transmission spectrum (see Sect. \ref{sec: resultsspectra}).

As this is not the case, our results suggest that the 3D NLTE models perform better than the others, specifically outperforming the commonly used 1D models, though further improvement can be achieved. For instance, recent studies have shown that magnetic fields impact limb darkening and the observed CLV in solar-type stars (e.g., \citealt{Ludwig2023}; \citealt{Kostogryz2024}).
Indeed, the presence of magnetic fields modifies the stellar atmospheric structure, which in turn affects the CLV of spectral lines. As can be seen from Fig. 2 and 3 in \citet{Ludwig2023}, both the averaged temperature profile and the spatial-temporal temperature fluctuation vary with magnetic field strength.
Specifically, the 3D radiation-magneto-hydrodynamic (RMHD) model atmospheres computed by \citet{Ludwig2023} were able to reduce, but not completely eliminate, the small systematic discrepancy between observations and models of the CLV.
Further work is required to investigate the role of stellar magnetic activity on transmission spectra and LCs of exoplanets. 
Further advancements in 3D NLTE synthetic spectra could be achieved from improvements to the model atmosphere, for example in terms of opacity binning or resolution, or from improvements to the model atom, such as more accurate collisional rates, or from lifting the trace-element assumption.

 
\subsection{CLV effects in the stellar grid}\label{sec: resultsgrid}
The intrinsic variations of stellar lines from the disk center to the limb are determined by the stellar parameters. The stellar effective temperature has the largest impact on the CLV effect (\citealt{Yan2015, Yan2017}).
Indeed, the widths of the \ion{Na}{I} D lines increase as the $T_\mathrm{eff}$ decreases. Line broadening in the wings occurs primarily because the ground state of \ion{Na}{I} is increasingly populated at low temperatures (i.e., the line opacity increases) (\citealt{Lind2011}). The CLV and $T_\mathrm{eff}$ dependence of the \ion{Na}{I} D lines is well illustrated in Fig. 3 in \citet{Czesla2015}.

In order to investigate the CLV effect across the stellar grid (Fig. \ref{fig: HRdiagram}) as a function of the stellar parameters of $T_\mathrm{eff}$, $\log g$ and [Fe/H], we compute the combined equivalent width ($W_\lambda$) of the \ion{Na}{I} D lines at different $\mu$-angles, by integrating the 3D NLTE synthetic spectra in the wavelength range 5866.0--5920.0\,\r{A}, that is the range of our synthetic spectra in the region of the \ion{Na}{I} D lines. We did not compute the $W_\lambda$ of the \ion{Na}{I} D$_1$ and D$_2$ separately because, in models with low $T_\mathrm{eff}$ (i.e. 4500\,K), the wings of the two lines become extremely broadened, causing them to blend together. 
We show the ratio of $W_\lambda$ at the limb ($\mu=0.1$) to that at the disk center ($\mu=1.0$) in Fig. \ref{fig: CLV effect}, with contour plots illustrating how this ratio varies with effective temperature and surface gravity at a sodium abundance of [Na/Fe] = 0.

The relative difference between the line strength at the limb and at the disk center shows significant variations across the stellar grid of 3D NLTE spectra, ranging between about $\pm 20$\%. The largest positive variations, meaning a stronger line at the limb, occur at high temperatures (i.e., $T_\mathrm{eff}=6500$\,K) and low surface gravities (i.e., $\log g=4.0$\,dex), where the line contrast reaches up to 28\% for [Fe/H]=$-0.5$. Conversely, the lines become weaker at the limb in the rest of the grid, particularly in metal-rich stars (i.e., [Fe/H]=+0.5) with low $T_\mathrm{eff}$. For instance, at $T_\mathrm{eff}=5000$\,K and $\log g = 5.0$\,dex, the difference decreases by as much as $-21$\%. 
This could be explained as a line-strength effect: the cooler the star, the stronger the lines, and the more they are weakened at the limb compared to disk center.

In 1D LTE (see Fig. \ref{fig: CLV effect in 1D} in Appendix) the lines are always weaker at the limb across the entire grid, with variations ranging from about $-38$\% to $-46$\% for lower metallicity. In 3D NLTE instead (see Fig. \ref{fig: CLV effect}), for high $T_\mathrm{eff}$, the lines at the limb become stronger than at the disk center, especially for low metallicity (i.e., [Fe/H]=$-0.5$).
This occurs due to the velocity fields that naturally arise in the 3D model atmosphere, which are absent in the 1D model. This effect can be reproduced in 1D by using, for instance, a depth-dependent $v_\mathrm{mic}$ \citep{Takeda2022}.


In addition to the CLV effect, we also observed that the difference in equivalent width between 3D and 1D models is as significant as the CLV effect itself across the entire grid. A more detailed discussion about this effect will be provided in a forthcoming paper, which will extend the grid to include lower metallicity models and giants. 


Also, Fig. \ref{fig: CLVEWsbestfit} shows the CLV of the \ion{Na}{I} D lines in the four host stars analyzed in this work, that is their $W_\lambda$ as a function of $\mu$-angles. These were obtained from the best-fit synthetic spectra in 1D LTE, 1D NLTE, 3D LTE, and 3D NLTE (see Sect. \ref{sec: synthspectra}). It is evident that the 1D LTE model dramatically exaggerates the CLV effect compared to the 3D NLTE model. The 1D models underestimate the equivalent width more toward the limb due to the lack of velocity fields, as previously shown by \citet{Reiners2023}. The LTE underestimation toward the limb suggests increased photon losses in shallower layers, as observed by \citet{Canocchi2024}.

\subsection{CLV+RM effects in the stellar grid}\label{sec: HD209stars}
This section illustrates how the CLV+RM model for the HD 209458b planetary system would change if the planet were orbiting around different types of FGK-type stars. Considering the planetary and orbital parameters of this system (see Table \ref{tab: systemparams1}), we compute the CLV+RM model for the transmission light curve in a 0.75\,\r{A} passband, using 3D NLTE synthetic spectra from the stellar grid with a solar sodium abundance, that is [Na/Fe] = 0.

A contour diagram of the mid-transit values for the CLV+RM curves in the 0.75\,\r{A} passband is shown in Fig. \ref{fig: HD209_LC75_grid}. 
We choose to plot the mid-transit value of the LC as it appears to be the datapoint with the largest variation among models (e.g., see Fig. \ref{fig: HD209results}). From the contour diagrams, it is evident that the amplitude of the CLV+RM curve increases with lower $T_\mathrm{eff}$ and higher $\log g$. The overall variations are within 
0.26\%. 
This result aligns with the findings of \citet{Yan2017}, according to whom stars with lower $T_\mathrm{eff}$ exhibit a stronger CLV+RM effect.
Nevertheless, the shape of the CLV+RM in transmission spectra and LCs is determined by orbital and planetary parameters, and might be important to correct for it also for stars with high $T_\mathrm{eff}$, as demonstrated in the case of HD 209458b. Indeed, from Fig. \ref{fig: HD209_LC75_grid}, it can be noticed that this planet lies on a part of the grid that is minimally affected by the CLV+RM effect, yet our analysis shows that accounting for it is essential when interpreting transmission spectra and LCs. If HD 209458b were orbiting a cooler star, the CLV+RM effect would be even more pronounced, making the correction even more critical.

Moreover, aside from the host star, several orbital parameters play a crucial role in shaping the CLV+RM model in transmission spectra and LCs. Among these, the impact parameter ($b$), the planet-to-star radius ratio ($R_\mathrm{p}/R_\star$), and the sky-projected obliquity ($\lambda$) are particularly significant.
Specifically, the impact parameter, which is related to the orbit inclination, and $R_\mathrm{p}/R_\star$ influence the amplitude of the curve (e.g., \citealt{Canocchi2024}). The actual shape is instead determined by $\lambda$, which describes the transit trajectory on the stellar disk. 


\section{Conclusions}\label{sec: conclusions}
In this work, we model the CLV of the strong \ion{Na}{I} D lines at 5890\,\r{A} and 5896\,\r{A} of the host stars of four giant planets using 3D RHD stellar models and NLTE line formation. We apply the synthetic spectra to compute the CLV+RM effect that affects the transmission spectra as well as the LCs of the four targets. We then correct the data taken with the high-resolution ESPRESSO spectrograph at the VLT during different campaigns for this effect. We also make a comparison with other synthetic spectra, and specifically, with the commonly used 1D LTE models computed with the \texttt{MARCS} (\citealt{Gustafsson2008}) grid, as well as models in 1D NLTE and 3D LTE. For the 3D simulations, for each model, we use five statistically independent snapshots from the recently extended and refined \textsc{Stagger}-grid, published in \citet{RodriguezDiaz2023}.

Our work strengthens the previous detections of \ion{Na}{I} in the WASP systems, improving the accuracy of the measurements and making the detections more statistically significant, except for systems in heavy misaligned orbit such as WASP-127b. We also confirm the findings of CB21, namely that there is no detection of \ion{Na}{I} in the atmosphere of HD 209458b at the layers probed by high-resolution spectroscopy. Indeed, the CLV+RM model computed using 3D NLTE synthetic spectra perfectly matches the observed transmission spectrum of the \ion{Na}{I} D lines, and it is in good agreement with the transmission LCs as well. 

For some configurations of the planet-star system, the CLV+RM effects do not significantly affect the resulting transmission spectra, as in the case of WASP-127b which is in an extremely misaligned orbit. However, for some other planets, like HD 209458b, the CLV+RM effect can be as strong as the planetary atmospheric feature and, depending on the orbital geometry, it can even overlap with the expected planetary atmosphere tracks (e.g., \citealt{Czesla2015}). For these planets, with the transmission spectra now achieving high-quality data due to advanced spectrographs such as ESPRESSO, the CLV+RM effect is within the signal-to-noise ratio of these observations and can no longer be disregarded. 
We conclude that 3D NLTE synthetic spectra can be used to improve the accuracy of the analyses of transmission data of exoplanets' atmosphere from current and future ground-based high-resolution spectrographs, such as the forthcoming ANDES on the Extremely Large Telescope (ELT; \citealt{Palle2023}). The correction of the CLV+RM effect will become even more crucial when studying colder planets, such as super-Earths, where the planetary atmosphere track will overlap with these stellar effects even more.
For this purpose, we publish a grid of 3D NLTE synthetic spectra for \ion{Na}{I}, for FGK-type stars with the following stellar parameters:
\begin{itemize}
    \item Effective temperature ($T_\mathrm{eff}$) between 4500 and 6500\,K, in steps of 500\,K;
    \item Surface gravity ($\log g$) between 4.0 and 5.0 in steps of 0.5;
    \item  Metallicity ([Fe/H]) between -0.5 and +0.5, in steps of 0.5;
    \item Sodium abundance ([Na/Fe]) between -0.5 and +0.5, in steps of 0.5. 
\end{itemize}
Together with the grid, we publish a Python interpolation routine that allows to compute intensity spectra for any given stellar parameters and for any given $\mu$-angles. In this way, the methodology presented in this work can be easily extended to other FGK host-stars in order to compute a more accurate CLV+RM effect of the \ion{Na}{I} D lines for any planet-star system.

An accurate analysis of the transmission spectra in the optical wavelength region from ground-based facilities is fundamental to complement the high-quality space-based observations in the infrared from telescopes such as the current JWST (\citealt{Gardner2006}; \citealt{Barstow2015}), and in the future from ARIEL (\citealt{Tinetti2016, Tinetti2018}).

Even though we do not detect any \ion{Na}{I} absorption in HD 209458b from our high-resolution transmission spectra, absorption from this species can still be observed at low- and medium-resolution. This has been recently shown by \citet{Carteret2024}, who investigated the impact of the RM effect on space-based data of HD 209458b. Their findings suggest that the observed absorption feature could be explained by absorption in the wings of the \ion{Na}{I} D lines, and not from the core probed at high-resolution.
They also found that absorption signatures from \ion{Na}{I} at medium resolution with instruments such as JWST/NIRSPEC or HST/STIS can be significantly biased by the RM effect if not properly accounted for. Therefore, an important issue to address in the near future would be to evaluate the impact of 3D NLTE synthetic stellar spectra on these types of datasets as well.

Finally, several other spectral lines relevant to transmission spectroscopy have been shown to be significantly affected by strong 3D NLTE effects (e.g., the H$\alpha$ line, \citealt{Amarsi2018}; the \ion{K}{I} resonance line, \citealt{Canocchi2024}). Future work includes the study and investigation of the impact of 3D NLTE models on other atomic species detected in exoplanets' atmosphere, such as \ion{Ca}{I}, \ion{Ca}{II} (e.g., \citealt{Yan2019}), \ion{Fe}{I} (e.g., \citealt{Hoeijmakers2018}), \ion{Fe}{II} (e.g., \citealt{Bello2022}), \ion{H}{I} Balmer lines (e.g., \citealt{Fossati2023}), and \ion{Mg}{I} triplet (e.g., \citealt{Hoeijmakers2020}; \citealt{Prinoth2024}). 
We refer the reader to \citet{LindAmarsi20242} for a review of 3D NLTE stellar models.

In a forthcoming paper, we will extend the grid of 3D NLTE spectra for \ion{Na}{I} to FGK-giants (i.e. $\log g=2.0-3.5$) and metal-poor stars, with metallicity down to [Fe/H]=$-4$, in order to explore the impact of the 3D model atmosphere on stellar abundance determination.


\begin{acknowledgements} 
We thank the anonymous referee for their comments, which have improved the quality of the manuscript.
GC and KL acknowledge funds from the Knut and Alice Wallenberg foundation. KL also acknowledges funds from the European Research Council (ERC) under the European Union’s Horizon 2020 research and innovation programme (Grant agreement No. 852977).
GM acknowledges financial support from the Severo Ochoa grant CEX2021-001131-S and from the Ram\'on y Cajal grant RYC2022-037854-I funded by MCIN/AEI/ 10.13039/501100011033 and FSE+.
We thank the PDC Center for High Performance Computing, KTH Royal Institute of Technology, Sweden, for providing access to computational resources and support. 
The computations were enabled by resources provided by the National Academic Infrastructure for Supercomputing in Sweden (NAISS), partially funded by the Swedish Research Council through grant agreement no. 2022-06725, at the PDC Center for High Performance Computing, KTH Royal Institute of Technology (project numbers NAISS 2023/1-15 and NAISS 2024/1-14).
This research has made use of NASA's Astrophysics Data System (ADS) bibliographic services. 
We acknowledge financial support from the Agencia Estatal de Investigaci\'on of the Ministerio de Ciencia e Innovaci\'on MCIN/AEI/10.13039/501100011033 and the ERDF “A way of making Europe” through project PID2021-125627OB-C32, and from the Centre of Excellence “Severo Ochoa” award to the Instituto de Astrofisica de Canarias.
We acknowledge the community efforts devoted to the development of the following open-source packages that were used in this work: numpy (\url{numpy.org}), matplotlib (\url{matplotlib.org}) and astropy (\url{astropy.org}). 

\end{acknowledgements}

\bibliographystyle{aa_url}
\bibliography{references} 


\appendix 
\section{Additional tables and figures}
The wind velocity in the three WASP planets is computed as the Doppler shift of the line centroids from the Gaussian fit ($\lambda_{fit}$) with respect to the rest frame wavelength ($\lambda_0$), as follows:
\begin{equation}
    v_{wind}= c \ \frac{\lambda_{fit} - \lambda_0}{\lambda_0} \ ,
    \label{eq: doppler}
\end{equation}
with $\lambda_0$ being the rest frame wavelength of the \ion{Na}{I} D lines. We first compute $v_{wind}$ separately for the \ion{Na}{I} D$_1$ and D$_2$, and then we take the weighted average of the two values, as reported in Table \ref{tab: winds}. The Table presents the values from the best-fit Gaussians after correcting for the CLV+RM effect using 3D NLTE stellar models. 
\begin{table}[H]
\centering 
\caption{Wind velocities, standard deviation ($\sigma$), and full width at half maximum (FWHM) of the Gaussian fits for the three gas giants analyzed in this work where \ion{Na}{I} was detected.}\label{tab: winds} 
\begin{tabular}{l c c c c} \hline \hline
\noalign{\smallskip}
Planet & Line & Velocity & $\sigma$ & FWHM \\
 & & (km\,s$^{-1}$) & (\r{A}) & (\r{A}) \\ \hline 
 \noalign{\smallskip}
WASP-52b & D$_1$ & $-5.3 \pm 0.8$ & $0.084 \pm 0.015$ & $0.19 \pm 0.03$ \\
 & D$_2$ & $-8.5 \pm 0.3$ & $0.030 \pm 0.006$ & $0.07 \pm 0.01$ \\
 & mean & $-8.0 \pm 0.3$ & $0.038 \pm 0.006$ & $0.09 \pm 0.01$ \\ 
WASP-76b & D$_1$ & $-4.8 \pm 0.7$ & $0.199 \pm 0.013$ & $0.47 \pm 0.03$ \\
 & D$_2$ & $-3.7 \pm 0.8$ & $0.245 \pm 0.015$ & $0.58 \pm 0.03$ \\
 & mean & $-4.3 \pm 0.5$ & $0.219 \pm 0.009$ & $0.52 \pm 0.02$ \\ 
WASP-127b & D$_1$ & $-3.7 \pm 0.9$ & $0.070 \pm 0.019$ & $0.17 \pm 0.04$ \\
 & D$_2$ & $-3.2 \pm 0.5$ & $0.102 \pm 0.009$ & $0.24 \pm 0.02$ \\
 & mean & $-3.2 \pm 0.4$ & $0.096 \pm 0.008$ & $ 0.23 \pm 0.02$ \\ 
\hline 
\end{tabular}\\
\end{table}
\begin{table}[H]
\centering 
\caption{Measured depths (\%) of the Gaussian fits to the \ion{Na}{I} D lines in the transmission spectra for the four targets analyzed in this work, after correcting for the CLV+RM effect with different stellar models (1D LTE, 1D NLTE, 3D LTE, and 3D NLTE).}\label{tab: Nadetmodels} 
\begin{tabular}{l c c c c} \hline \hline
\noalign{\smallskip}
Planet & Model & D$_1$ & D$_2$ & $\sigma$ \\ 
 & & (\%) & (\%) & (D$_1$/D$_2$) \\ \hline
 \noalign{\smallskip}
HD 209458b & 1D LTE & $0.18 \pm 0.02$ & $0.13 \pm 0.02$ & 6.9/6.5\\ 
 & 1D NLTE & $0.14 \pm 0.02$ & $0.16 \pm 0.01$ & 5.3/6.2\\ 
 & 3D LTE & $0.16 \pm 0.01$ & $0.16 \pm 0.01$ & 6.1/6.2\\
 & 3D NLTE & \multicolumn{2}{c}{$0.000 \pm 0.026$} & - \\
WASP-52b & 1D LTE & $1.16 \pm 0.19$ & $1.69 \pm 0.32$ & 6.1/5.3 \\
 & 1D NLTE & $1.22 \pm 0.19$ & $1.71 \pm 0.32$  & 6.4/5.3 \\
 & 3D LTE & $1.17 \pm 0.19$ & $1.69 \pm 0.32$ & 6.1/5.3 \\
 & 3D NLTE & $1.25 \pm 0.19$ & $1.75 \pm 0.32$ & 6.7/5.5 \\
WASP-76b & 1D LTE & $0.384 \pm 0.020$ & $0.423 \pm 0.021$ & 10.1/11.1   \\ 
 & 1D NLTE & $0.377 \pm 0.020$ & $0.415 \pm 0.021$ & 9.9/10.9 \\ 
 & 3D LTE & $0.366 \pm 0.021$ & $0.402 \pm 0.021$ & 9.6/10.6  \\ 
 & 3D NLTE & $0.376 \pm 0.021$ & $0.389 \pm 0.022$ & 9.9/10.2\\ 
WASP-127b & 1D LTE & $0.25 \pm 0.05$ & $0.54 \pm 0.04$ & 5.4/13.7 \\
 & 1D NLTE & $0.21 \pm 0.05$ & $0.51 \pm 0.04$ & 4.3/12.5 \\
 & 3D LTE & $0.22 \pm 0.05$ & $0.53 \pm 0.04$ & 4.7/13.1 \\
 & 3D NLTE & $0.21 \pm 0.05$ & $0.51 \pm 0.04$ & 4.4/12.7\\
\hline 
\end{tabular}\\
\begin{flushleft}
\textbf{Notes:}\\
In the last column, the significance of the detections is shown.
\end{flushleft}
\end{table}
In Table \ref{tab: Nadetmodels}, the absorption depths of the \ion{Na}{I} D$_1$ and D$_2$ lines, corrected for the CLV+RM effect using different stellar models, are shown.

It is interesting to note that for HD 209458b, all models except the 3D NLTE model detect \ion{Na}{I} with a significance level larger than $5\sigma$. The only non-detection occurs when using the 3D NLTE stellar spectra. 

For WASP-52b, a system with an aligned orbit, the 3D NLTE model results in a detection with a slightly higher confidence level compared to other models, although the measured depths are consistent within their uncertainties.

In the cases of the heavily misaligned systems WASP-76b and WASP-127b, the use of a 3D NLTE model does not significantly enhance the detection's significance. The measured absorption depths from various models agree well within their uncertainties.

\begin{table*}
\centering 
\caption{Parameters of the 3D NLTE synthetic spectra computed in the grid for this work.}\label{tab: synthspectraparams} 
\begin{tabular}{l c c c} \hline \hline
\noalign{\smallskip}
$\mathrm{[Fe/H]}$ & $\log{g}$ & $T_\mathrm{eff}$ & $A$(Na)$^{(a)}$ \\ 
(dex) &  (cm s$^{-2}$) & (K) & (dex) \\ \hline 
 \noalign{\smallskip}
$-0.5$ & 4.0 & 4500, 5500, 6000, 6500  & 5.22, 5.72, 6.22 \\ 
 & 4.5 & 4500, 5000, 5500, 6000, 6500 & 5.22, 5.72, 6.22 \\
 & 5.0 & 4500, 5000, 5500 & 5.22, 5.72, 6.22 \\
$0.0$ & 4.0 & 4500, 5000, 5500, 6000, 6500 & 5.72, 6.22, 6.72\\
 & 4.5 & 4500, 5000, 5500, 6000, 6500 & 5.72, 6.22, 6.72 \\
 & 5.0 & 4500, 5000, 5500 & 5.72, 6.22, 6.72\\
$+0.5$ & 4.0 & 4500, 5000, 5500, 6000, 6500 & 6.22, 6.72, 7.22\\
 & 4.5 & 4500, 5000, 5500, 6000, 6500 & 6.22, 6.72, 7.22 \\
 & 5.0 & 4500, 5000, 5500 & 6.22, 6.72, 7.22\\ 
\hline 
\end{tabular}\\
\begin{flushleft}
\textbf{Notes:}\\
$^{(a)}$ We take as the reference for the solar abundance the value reported in \citet{Asplund2021}: $A\mathrm{(Na)}_\odot = 6.22 \pm 0.03$. In the other abundance notation, this corresponds to [Na/Fe]=0.0 for the solar metallicity models (i.e. [Fe/H]=0.0).

\end{flushleft}
\end{table*}
\begin{figure*}
\resizebox{\hsize}{!}
          {\includegraphics[width=1.0\textwidth]{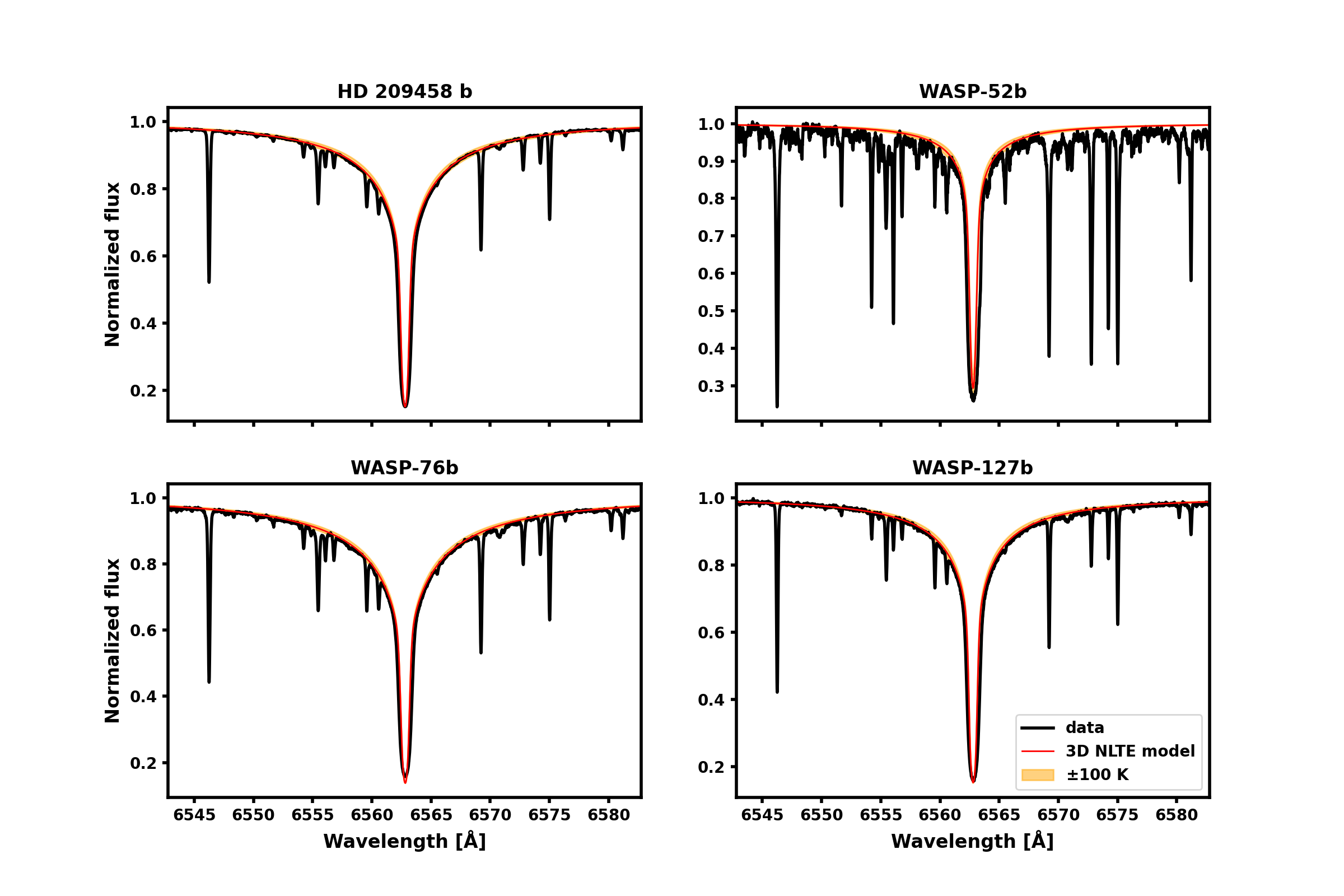}}
          \caption{Master Out around the H$\alpha$ line profiles observed with ESPRESSO in the four target stars, compared to the 3D NLTE model with $T_\mathrm{eff}$, $\log g$, [Fe/H], and $v\sin{i_\star}$ listed in Table \ref{tab: systemparams1}. The orange shaded region indicates the effect of adjusting the $T_\mathrm{eff}$ by $\pm 100$\,K.}
          \label{fig: Halpha}
\end{figure*}
\begin{figure*}
   \centering
   \includegraphics[width=0.49\textwidth]{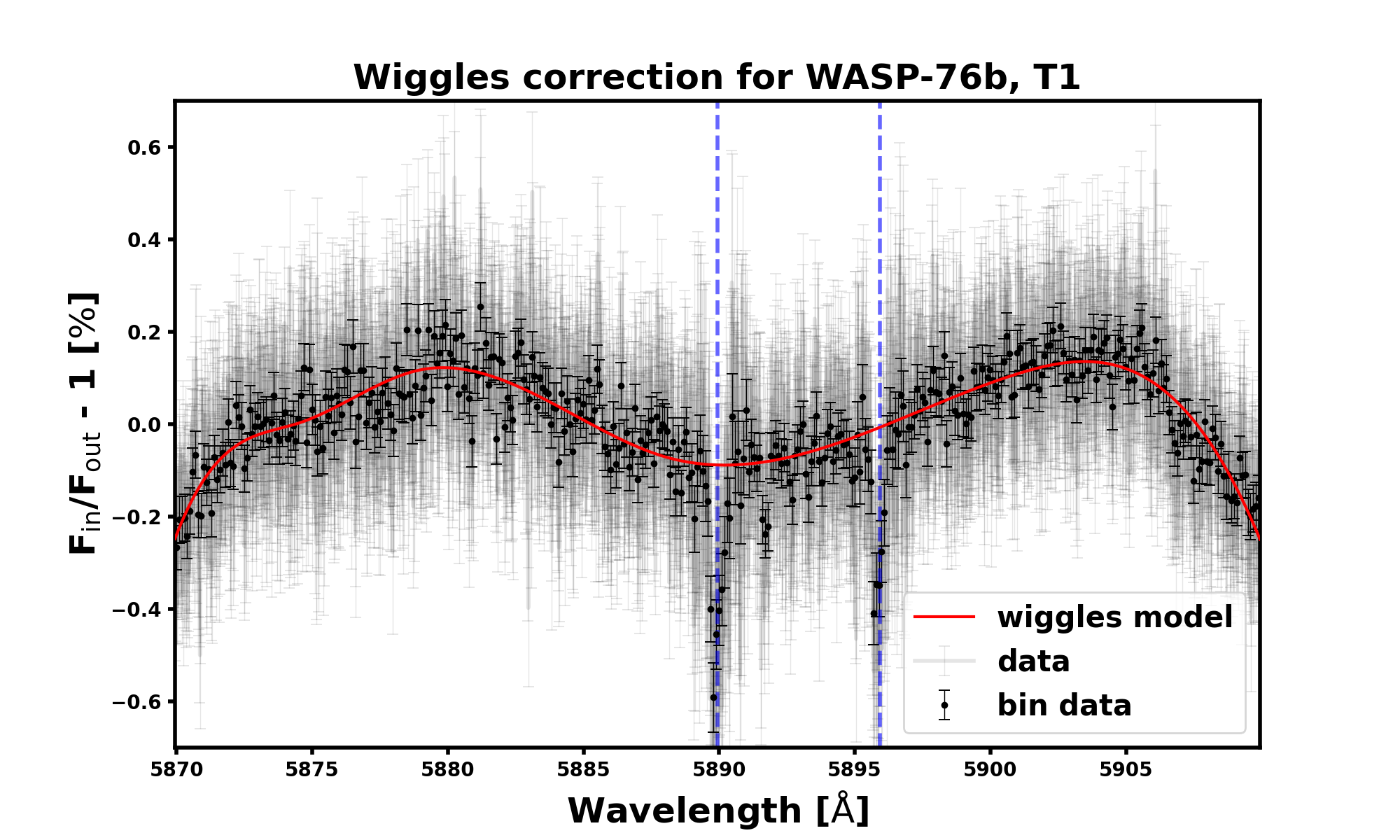}
   \includegraphics[width=0.49\textwidth]{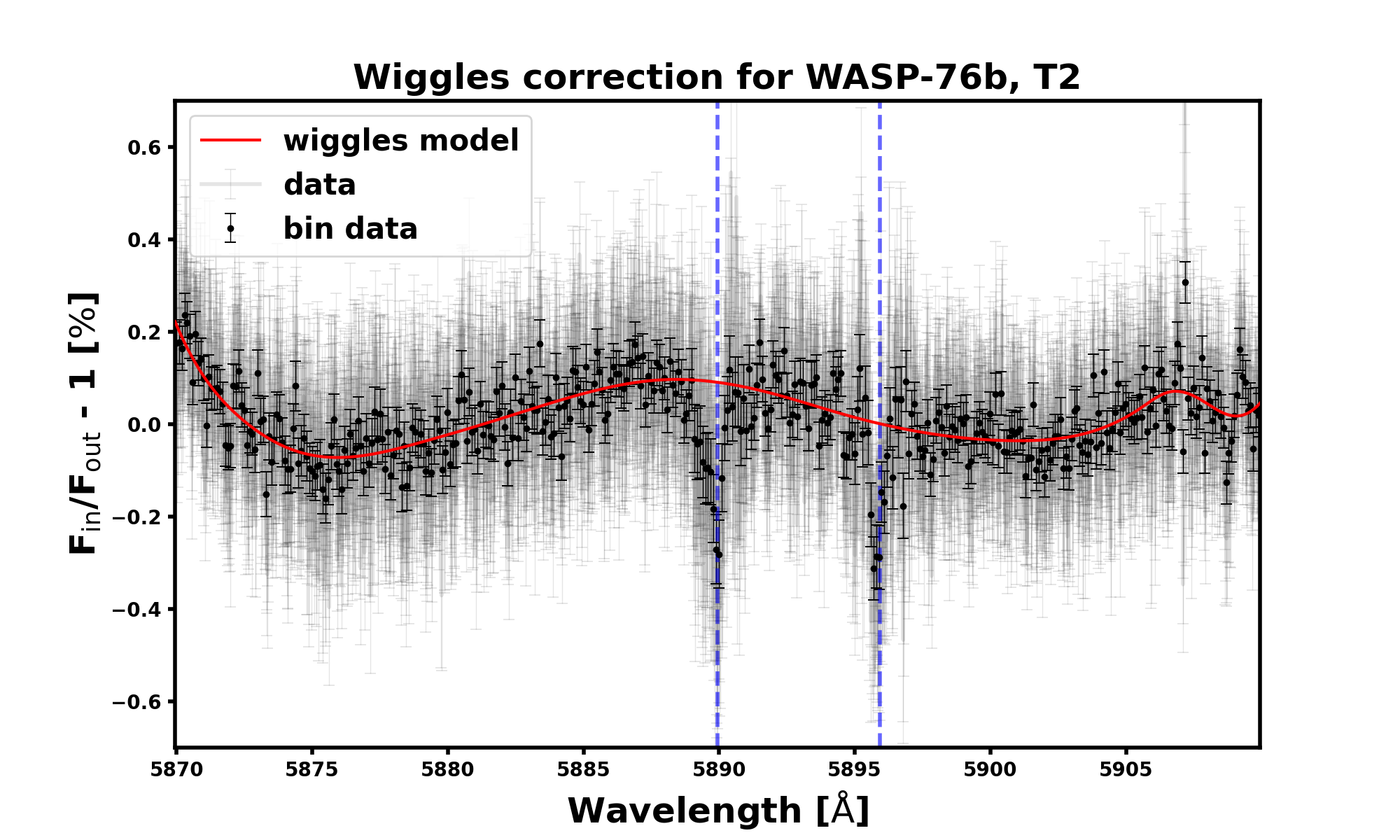}
   \caption{Transmission spectra (in black) of WASP-76b with overplotted the wiggles model, that is a third-order spline (in red). \textit{Left}: in- to out-of-transit flux observed by ESPRESSO on 2018-09-03. \textit{Right}: in- to out-of-transit flux observed by ESPRESSO on 2018-10-31.}
   \label{fig: wiggleexample}
\end{figure*}
\begin{figure*}
   \resizebox{\hsize}{!}
            {\includegraphics[width=10cm]{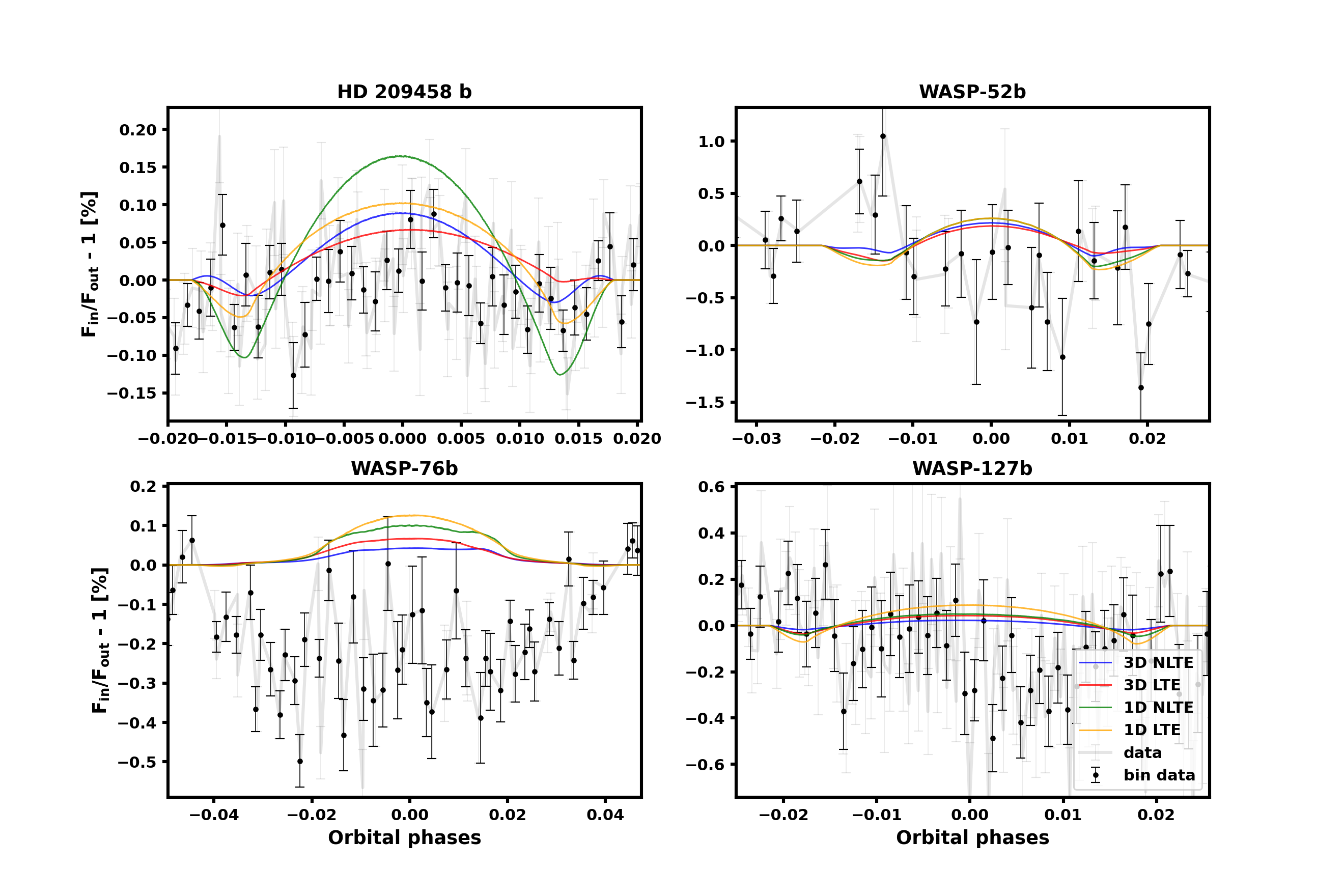}}
            
      \caption{Transmission LCs in a 0.75\,\r{A} passband centered on the \ion{Na}{I} D lines. The gray dots represent the original data, whereas the black dots are the data binned to 0.001 in orbital phase. The CLV+RM model from different synthetic stellar spectra is overplotted, colored as in the figure's legend. It is worth noticing the different scale of the y-axis.} 
              
         \label{fig: transLCs075}
\end{figure*}
\begin{figure*}
   \resizebox{\hsize}{!}
            {\includegraphics[width=10cm]{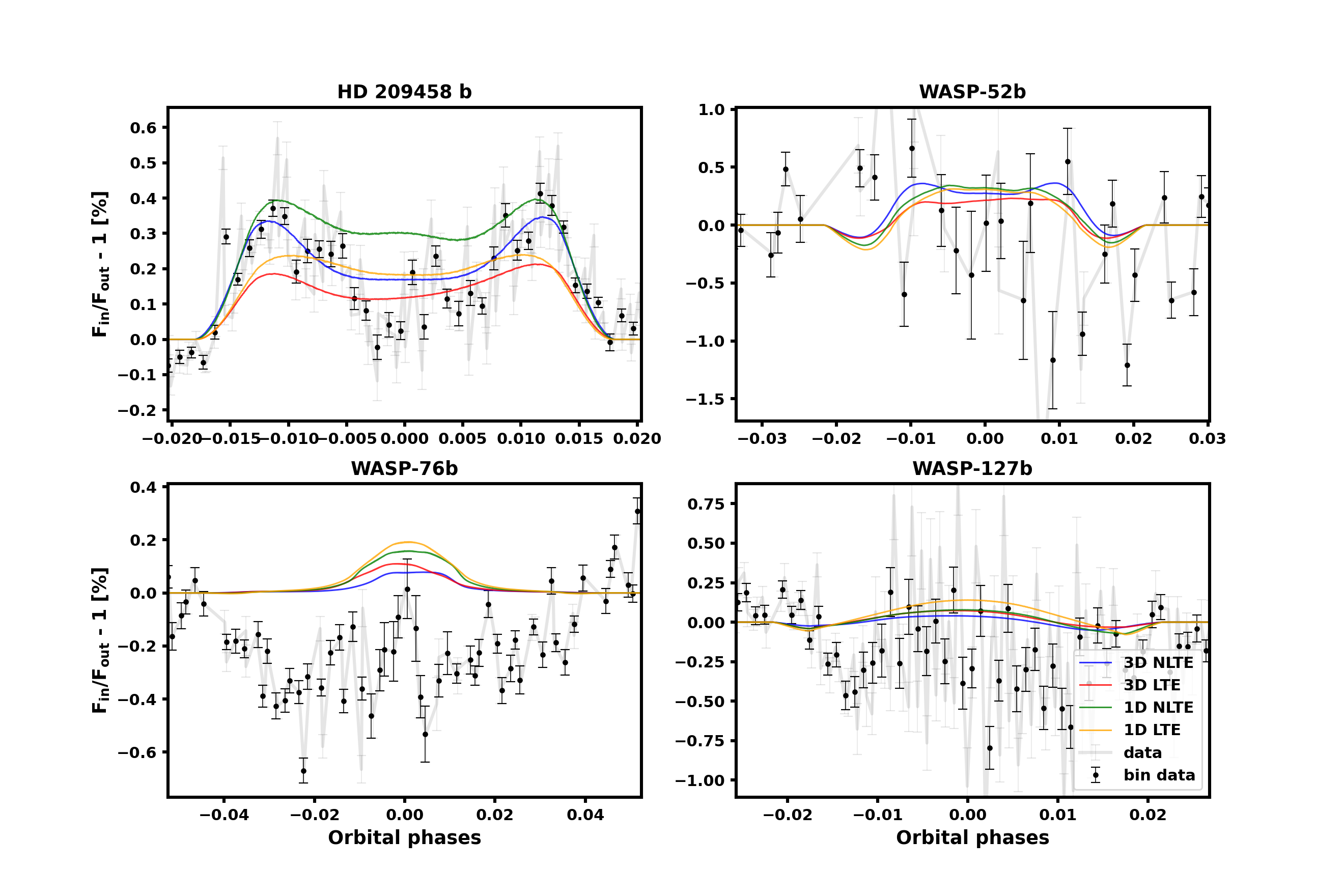}}
            
      \caption{Transmission LCs in a 0.4\,\r{A} passband centered on the \ion{Na}{I} D lines. The gray dots represent the original data, whereas the black dots are the data binned to 0.001 in orbital phase. The CLV+RM model from different synthetic stellar spectra is overplotted, colored as in the figure's legend. It is worth noticing the different scale of the y-axis.} 
              
         \label{fig: transLCs04}
\end{figure*}
\begin{figure*}
    \resizebox{\hsize}{!}
            {\includegraphics[width=15cm]{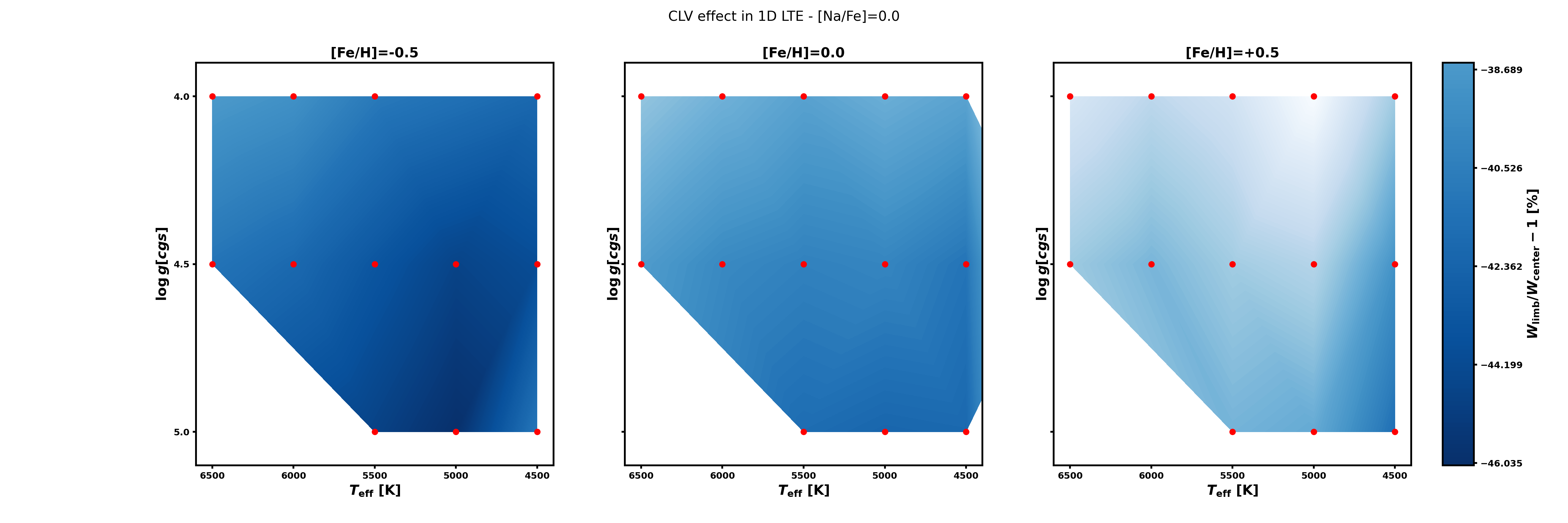}}
            
      \caption{The CLV effect across the stellar grid for FGK-type stars. The panels show the ratio of the combined equivalent width of the \ion{Na}{I} D lines from 1D LTE synthetic spectra at the limb ($\mu=0.1$) and at the disk center ($\mu=1.0$). The red dots indicate the grid nodes where the synthetic spectra have been computed.} 
         \label{fig: CLV effect in 1D}
\end{figure*}

\end{document}